\documentclass[%
 amsmath,amssymb,
 aps,
 prd,
twocolumn, superscriptaddress
]{revtex4-1}

\usepackage[normalem]{ulem}

\usepackage{graphicx}
\usepackage{dcolumn}
\usepackage{bm}
\usepackage{hyperref}
\usepackage{amssymb}
\usepackage{xcolor}

\usepackage[dvipsnames]{xcolor}
\hypersetup{colorlinks,citecolor=OrangeRed}
\hypersetup{colorlinks=true, linkcolor=blue, urlcolor=MidnightBlue}

\begin{document}

\preprint{APS/123-QED}

\title{White dwarf structure in $f(R,T,L_m)$ gravity: beyond the Chandrasekhar mass limit}

\author{Edson Otoniel}
 \email{edson.otoniel@ufca.edu.br}
 \affiliation{Instituto de Formaç\~ao de Educadores, Universidade Federal do Cariri, R. Oleg\'ario Emidio de Araujo, s/n – Aldeota, 63260-000 Brejo Santo, CE, Brazil
}

\author{Juan M. Z. Pretel}
 \email{juan04manuel91@gmail.com}
 \affiliation{Centro Brasileiro de Pesquisas F{\'i}sicas, Rua Dr.~Xavier Sigaud, 150 URCA, Rio de Janeiro CEP 22290-180, RJ, Brazil
}

\author{Cl{\'e}sio E. Mota}
 \email{clesio200915@hotmail.com}
 \affiliation{Departamento de F\'isica, CFM - Universidade Federal de Santa Catarina; C.P. 476, CEP 88.040-900, Florian\'opolis, SC, Brasil.
 }

\author{César O. V. Flores}
 \email{cesarovfsky@gmail.com}
 \affiliation{Centro de Ci\^encias Exatas, Naturais e Tecnol\'ogicas, CCENT - Universidade Estadual da Regi\~ao Tocantina do Maranh\~ao; C.P. 1300,\\ CEP 65901-480, Imperatriz, MA, Brasil
}
 \affiliation{Departamento de F\'isica, CCET - Universidade Federal do Maranh\~ao, Campus Universit\'ario do Bacanga; CEP 65080-805, S\~ao Lu\'is, MA, Brasil
}

\author{Victor B. T. Alves}
 \email{victor.bruno@discente.ufma.br}
 \affiliation{Departamento de F\'isica, CCET - Universidade Federal do Maranh\~ao, Campus Universit\'ario do Bacanga; CEP 65080-805, S\~ao Lu\'is, MA, Brasil
}

\author{Franciele M. da Silva} 
\email{franciele.m.s@ufsc.br}
\affiliation{Departamento de F\'isica, CFM - Universidade Federal de Santa Catarina, \\ Caixa Postal 5064, CEP 880.35-972, Florian\'opolis, SC, Brazil.}
\affiliation{Theoretical Astrophysics, Institute for Astronomy and Astrophysics, University of T\"{u}bingen, 72076 T\"{u}bingen, Germany}

\date{\today}

\begin{abstract}
In this work, we investigate the relativistic structure of white dwarfs (WDs) within the framework of modified gravity theory $f(R, T, L_m) = R + \alpha T L_m$, which introduces a non-minimal coupling between matter and curvature. Using a realistic equation of state (EoS) that includes contributions from a relativistic degenerate electron gas and ionic lattice effects, we solve the modified Tolman-Oppenheimer-Volkoff (TOV) equations for two standard choices of the matter Lagrangian density: $L_m = p$ and $L_m = -\rho$. We show that the extra $\alpha TL_m$ term significantly alters the mass-radius relation of WDs, especially at high central densities $( \rho_c \gtrsim 10^8 - 10^9\,\rm  g/cm^3)$, allowing for stable super-Chandrasekhar configurations. In particular, depending on the sign and magnitude of the parameter $\alpha$, the maximum mass can increase or decrease, and in some regimes, the usual critical point indicating the transition from stability to instability disappears. Our findings suggest that $f(R,T,L_m)$ gravity provides a viable framework to explain the existence of massive WDs beyond the classical Chandrasekhar limit. Using Bayesian inference with WD observational data, we further constrain the coupling parameter $\alpha$ for the two choices of the Lagrangian density $L_m$.
\end{abstract}

\maketitle


\section{Introduction}

In addition to advancing our understanding of many aspects of the universe, the theory of gravity proposed by Einstein over a century ago has undergone, and continues to undergo, a substantial number of experimental tests. Among these tests are the precession of Mercury’s perihelion \cite{1915SPAW831E}, predicted with remarkable precision; the recent detections of gravitational waves generated by binary black hole systems \cite{GWblackhole} and neutron star (NS) mergers \cite{PhysRevLett.119.161101}, observed by LIGO-Virgo collaborations; and the first image of a black hole shadow obtained by the \textit{Event Horizon Telescope} project \cite{2019ApJ...875L...1E}.

Nevertheless, there are some phenomena at the larger (cosmological) scale that are not well described within the context of General Relativity (GR). For example, GR fails to explain the accelerated expansion of the universe without further refinement. To overcome this limitation, two possibilities were proposed. One is that there is a large amount of dark energy with negative pressure in the universe. The other is that the predictions of GR may be biased on the cosmological scale, so there are alternative theories of gravity \cite{PRD03sergei, Allemandi, Koyama2016, repsergei, Shankaranarayanan2022}, some of which extend GR through the introduction of additional terms in the standard Einstein-Hilbert action, such as massive gravity \cite{deRham:2010kj}, Brans-Dicke gravity \cite{Brans:1961sx}, $f(R)$ gravity \cite{1970MNRAS.150....1B, DeFelice:2010aj, Capozziello:2011et, Nojiri:2010wj, fder1} and other extensions \cite{Harko2010EPJC, Harko2011PRD, Harko:2010mv, Odintsov:2013iba, HARKO2021100886}. 

Additionally, Haghani and Harko proposed a generalized theory of gravity known as \( f(R, T, L_m) \) gravity \cite{Haghani2021}, in which the gravitational Lagrangian density is assumed to be an arbitrary function of the Ricci scalar \( R \), the trace of the energy-momentum tensor \( T \), and the matter Lagrangian density \( L_m \)~\cite{Haghani2013}. This approach unifies the \( f(R, T) \)~\cite{Harko2011PRD} and \( f(R, L_m) \)~\cite{Harko2010EPJC} models and introduces a non-minimal coupling between matter and curvature. A particularly interesting and simple case is the gravity model given by the function \( f(R, T, L_m) = R + \alpha T L_m \), where \( \alpha \) is a free parameter controlling the strength of the matter-geometry coupling. In recent years, this theory has gained attention as a possible framework to explain astrophysical phenomena beyond GR, including the existence of compact stars with masses above traditional limits. Some studies have analyzed the behavior of NSs and quark stars (QSs) in this context, revealing that the choice of \( L_m \) (typically \( L_m = p \) or \( L_m = -\rho \)) significantly affects the stellar structure and the resulting mass-radius relations \cite{Mota2024, Pretel2024PS, Tangphati2025}. It is worth mentioning that \( f(R, T, L_m) \) gravity has also recently been used to describe anisotropic spherical configurations under the influence of an electric charge in Refs.~\cite{NASEER2025101840, NASEER2025101958}.

Regarding theoretical consistency, it is important to stress that, for the specific gravity model adopted here, $f(R,T,L_m)=R+\alpha T L_m$, one has $f_R\equiv \partial f/\partial R = 1$ and $f_{RR}= 0$. Hence, the gravitational sector does not introduce higher-derivative corrections with an extra propagating scalar degree of freedom of the type present in generic \(f(R)\) models, and the usual ``ghost'' issue associated with an incorrect sign of \(f_R\) does not arise in the same way. Instead, physical viability in the stellar context is controlled by the regularity of the modified TOV system and by the absence of singular effective couplings in the denominators of the structure equations; therefore, in this work we restrict \(\alpha\) to the intervals for which the solutions are well-defined and smoothly recover GR when \(\alpha\to 0\). It is also noteworthy that Haghani and Harko \cite{Haghani2021} have investigated in detail the Dolgov-Kawasaki instability for the $f(R,T,L_m)$ theory, and they have obtained the criterion for the stability of the theory, $f_{RR}(0) \geq 0$. This corresponds to the standard stability condition found in $f(R)$ gravity \cite{fder1}, ensuring perturbative stability and the absence of tachyonic instabilities.

WDs are highly dense celestial objects resulting from the gravitational collapse of low- and intermediate-mass stars after nuclear fuel depletion in their cores. In other words, WDs are dense, hot remnants that cool over time. The study of these stars has significantly advanced our understanding of stellar evolution and compact object physics. Recent research has focused on the relation between WD mass and radius, revealing the Chandrasekhar mass limit as an upper bound \cite{Chandrasekhar:1931ih, Chandrasekhar:1935zz}. Notably, observations of peculiar over-luminous type-Ia supernovae such as SN 2007if, SN 2006gz, SN 2003fg, and SN 2009dc have suggested the existence of WDs with masses ranging from $2.1\, M_\odot$ to $2.8\, M_\odot$ (where $M_\odot$ is solar mass) \cite{SNLS:2006ics, Scalzo:2010xd, Hicken:2007ap, Yamanaka:2009dp, Silverman:2010bh, Taubenberger:2010qv}. Consequently, to understand the formation of super-Chandrasekhar mass WDs, extensive investigations have explored different contexts such as including super-strong uniform magnetized WDs \cite{Das:2012ai, Das:2013gd, Das:2014ssa, Franzon:2015gda, Deb:2021gwn}, WDs in modified gravity theories \cite{Das:2014rca, Jing_2016, Banerjee:2017uwz, Carvalho:2017pgk, Kalita:2018ldn, EslamPanah:2018evk, Liu:2018jhd,Rocha:2019nze, Wojnar:2020ckw, Kalita:2019yaj, Sarmah2022PRD, Kalita:2022tyx, Kalita2023, Li2024, PRETEL2025}, electrically charged WDs \cite{Liu:2014jna, Carvalho:2018kbt}, and rotating WDs \cite{Boshkayev:2011aet, Boshkayev:2012bq}. 

In view of the possible violation of the canonical Chandrasekhar mass limit, this work examines the relativistic structure and stability of WDs within the framework of modified \( f(R,T,L_m) \) gravity by adopting the functional form \( f(R,T,L_m) = R + \alpha T L_m \), which introduces a non-minimal coupling between matter and curvature. Employing a realistic EoS for the microphysics of the star and solving the corresponding modified TOV equations, we demonstrate that such theory allows for the existence of stable WD configurations with masses exceeding the classical Chandrasekhar limit. Beyond establishing the existence of super-Chandrasekhar branches, the novelty of the present framework is that the stellar structure becomes explicitly sensitive to the choice of \(L_m\), which acts as a ``discriminator'' for non-minimal matter--geometry couplings: for the same microphysical EoS, the two standard prescriptions \(L_m=p\) and \(L_m=-\rho\) yield qualitatively distinct trends for the maximum mass and compactness as functions of \(\alpha\). This feature provides an avenue to distinguish this model from other modified-gravity proposals that can mimic similar mass--radius curves at the background level: combining precise WD mass--radius measurements with independent information on the microphysics (composition, temperature and magnetic effects) can, in principle, break degeneracies and favor (or rule out) specific couplings through their characteristic \(L_m\)-dependence. These findings support the theoretical possibility of super-Chandrasekhar WDs within this gravitational setting, which may be relevant in the observational context of understanding the origin of peculiar over-luminous type Ia supernovae reported in the literature.

This article is organized as follows: In Sec.~\ref{sec2}, the composition of WD matter is discussed, including contributions from the relativistic degenerate electron gas and the ionic lattice. In Sec.~\ref{sec3}, we address the inverse beta-decay instability that may arise at high energy densities. Sec.~\ref{sec4} presents the modified TOV equations derived in the framework of \( f(R, T, L_m) \) gravity for two standard choices of the matter Lagrangian density: \( L_m = p \) and \( L_m = -\rho \). Sec.~\ref{sec5} is devoted to the numerical solutions of the stellar structure equations, where we analyze the mass-radius relations and compactness profiles for WDs under different values of the coupling parameter \( \alpha \). In the same section, a Bayesian analysis is carried out to identify the values of $\alpha$ that best fit the WD observational data. Finally, in Sec.~\ref{sec6}, we summarize our main results and discuss potential implications for the existence of super-Chandrasekhar WDs within this modified gravity context.


\section{White Dwarf Matter Composition}\label{sec2} 

As established in the foundational studies by ~\cite{salpeter_energy_1961, hamada_models_1961}, white dwarf (WD) matter predominantly consists of atomic nuclei embedded in a fully degenerate electron gas. In accordance with the approach of Otoniel \emph{et al.}~(2019)~\cite{Otoniel_2019}, the EoS for this matter is derived using updated atomic mass evaluations~\citep[see][and references therein]{wang_ame2012_2012,audi_ame2012_2012}. In the present context, we neglect the effects of magnetic fields on the WD matter EoS. The internal pressure within WDs arises primarily from degenerate electrons and the ionic lattice~\citep[see also][for lattice structures in the NS crust]{shapiro_black_2008}. Thus, by applying this formalism, the total pressure in WD matter, accounting for contributions from both the degenerate electron gas and the ionic lattice, is given by
\begin{equation}
p\left(k_F\right)=\frac{1}{3 \pi^2 h^3} \int_0^{k_F} \frac{k^4}{\sqrt{k^2+m_{\varepsilon}^2}} d k + p_L(Z) .
\end{equation}

The first term represents the pressure exerted by the relativistic degenerate electron gas. The integral arises from the momentum distribution of electrons up to the Fermi momentum $k_F$, which defines the maximum occupied momentum state at zero temperature. The quantity $k$ denotes the electron momentum, and the integrand accounts for the relativistic dispersion relation of the electrons. The effective electron mass is denoted by $m_{\varepsilon} = m_e c$, where $m_e$ is the rest mass of the electron and $c$ is the speed of light. The presence of $m_{\varepsilon}$ in the denominator ensures that the relativistic effects are included in the pressure calculation, which is crucial for modeling high-density environments such as the cores of WDs. The factor $1/(3 \pi^2 h^3)$ originates from the normalization of the momentum space volume in three dimensions, where $h$ is Planck's constant. This prefactor ensures the proper dimensional consistency and normalization of the integral over electron momenta. The second term, the pressure contribution from the ionic lattice, $p_L(Z)$, is given by the following expression:
\begin{equation}
p_L(Z) = \frac{1}{3} C e^2 n_e^{4/3} Z^{2/3} .
\end{equation}
This term describes the electrostatic pressure resulting from the Coulomb interactions among ions arranged in a crystalline lattice, typically assumed to form a body-centered cubic (bcc) structure in WD interiors. In this equation, $C$ is a dimensionless numerical constant associated with the lattice geometry, with a typical value of $C = -1.444$ for a bcc configuration. The negative sign reflects the binding nature of the electrostatic potential energy in the lattice. The term $e^2$ thus corresponds to the Coulomb interaction strength between electrically charged particles. The quantity $n_e$ is the number density of electrons, which, in the context of fully ionized WD matter, is directly related to the density of positive ions due to charge neutrality. $Z$ is the atomic number of an element, i.e., the number of protons in the nucleus of each atom of that element. This lattice pressure term is crucial for accurately modeling the total pressure in WD interiors, particularly at lower densities where the ionic contribution is non-negligible compared to the electron degeneracy pressure. Together, these two components describe the EoS for WD matter in the absence of magnetic fields, capturing both quantum degeneracy effects and the structural influence of the ion lattice.

The total energy density $\epsilon(k_F)$ of WD matter, incorporating contributions from nuclei, electrons, and the ionic lattice, is given by~\cite{chamel_stability_2013} as
\begin{align}
\epsilon\left(k_F\right) =&\ \epsilon_i+\epsilon_e+ \epsilon_L -\epsilon_\epsilon  \nonumber \\
=&\ n_i M(Z,A) c^2+\frac{1}{\pi^2 h^3} \int_0^{k_F} \sqrt{k^2+m_e^2} k^2 dk  \nonumber \\
& + C e^2 n_e^{4/3} Z^{2/3} - n_e m_e c^2 .
\end{align}
The first contribution, $\epsilon_i$, represents the rest-mass energy density of the fully ionized atomic nuclei, where $n_i$ is the number density of ions and $M(Z,A)$ is the nuclear mass of an ion with atomic number $Z$ and mass number $A$. In this work, we consider carbon as the constituent element of WD matter, adopting $Z = 6$ and $A = 12$, which corresponds to fully ionized ${}^{12}\mathrm{C}$. The nuclear mass $M(6,12)$ used in our calculations is obtained from experimental atomic mass evaluations, ensuring consistency with the most recent empirical data \citep[see][]{wang_ame2012_2012,audi_ame2012_2012}.

The second term, $\epsilon_e$, corresponds to the energy density of the degenerate electron gas, integrating the relativistic energy of electrons from zero momentum up to the Fermi momentum $k_F$. The integrand $\sqrt{k^2 + m_e^2} k^2$ incorporates both kinetic and rest energy of the electrons within the Fermi sea. The third contribution, $\epsilon_L = C e^2 n_e^{4/3} Z^{2/3}$, accounts for the energy density of the Coulomb lattice of ions and has been previously defined in the context of lattice pressure. The final term, $- n_e m_e c^2$, subtracts the rest-mass energy of the electrons, which is already implicitly included in the nuclear mass $M(Z,A)$. This correction avoids double counting the electron rest energy when computing the total energy density. Altogether, this formulation provides a consistent and comprehensive expression for the total energy density in WD matter under the assumption of a fully degenerate, magnetically unperturbed, and crystallized plasma.

\section{Inverse $\beta$-decay Reaction in White Dwarfs}\label{sec3}

At sufficiently high densities within the interiors of WDs, inverse $\beta$-decay reactions become energetically favorable, potentially leading to dynamical instabilities. This phenomenon, first proposed by Gamow in 1939~\cite{gamow_physical_1939} and later detailed in~\citep{shapiro_black_2008}, involves the electron capture process:
\[
A(N,Z) + e^- \rightarrow A(N+1, Z-1) + \nu_e .
\]
Such reactions reduce the number of electrons (responsible for generating the degeneracy pressure that supports the star against gravitational collapse) thereby softening the EoS. As a result, atomic nuclei become increasingly neutron-rich, decreasing both the electron energy density and pressure, which may ultimately drive the WD toward collapse.

The treatment of inverse $\beta$-decay processes in WDs relies on a thermodynamic formulation. From the relation $\epsilon_e + p_e = n_e \mu_e$, one can derive the Gibbs free energy per nucleon as
\begin{equation}
g(A,Z) = m_nc^2 + \frac{M(Z,A)c^2}{A} + \gamma_e\left[\mu_e - m_e c^2 + \frac{4}{3}\frac{\epsilon_L}{n_e} \right] ,
\label{secA:eq1}
\end{equation}
where $\gamma_e = Z/A$ denotes the proton-to-nucleon ratio, $m_n$ is the neutron mass, $M(Z,A)$ the nuclear mass, $\mu_e$ the electron chemical potential, and $\epsilon_L$ the lattice energy density.

The onset of inverse $\beta$-decay is expected to occur when the Gibbs free energy of the daughter nucleus becomes lower than that of the parent nucleus, satisfying the condition~\citep{chamel_maximum_2014}:
\begin{equation}
g(A,Z) \geq g(A,Z-1) .
\label{secA:gibb}
\end{equation}
Substituting Eq.~(\ref{secA:eq1}) into the inequality above yields~\cite{PhysRevD.92.023008}:
\begin{equation}
\mu_e + C e^2 n_e^{1/3} f(Z,Z-1) \geq \mu_e^\beta ,
\label{sec4:ineq2}
\end{equation}
where $\mu_e^\beta$ is the threshold electron chemical potential defined by the nuclear mass difference:
\begin{equation}
\mu_e^\beta(A,Z) \equiv M(Z-1,A)c^2 - M(Z,A)c^2 + m_e c^2 .
\end{equation}
Moreover, the function $f(Z,Z-1)$ encodes the Coulomb correction arising from the lattice structure and is given by
\begin{equation}
f(Z,Z-1) = Z^{5/3} - (Z-1)^{5/3} + \frac{1}{3} Z^{2/3} .
\end{equation}

To express the electron number density $n_e$ and the mass density $\rho$ of the electron gas, we use
\begin{eqnarray}
n_e &=& \frac{k_F^3}{3 \pi^2 \hbar^3} , \label{n_e} \\ 
\rho &=& \frac{1}{\gamma_e} m n_e ,
\end{eqnarray}
where $k_F$ is the electron Fermi momentum. From these expressions, the Fermi momentum can be written as:
\begin{equation}
k_F = \hbar \left( \frac{3 \pi^2 \rho}{m_n} \frac{Z}{A} \right)^{1/3} .
\end{equation}

Since the momentum of the nuclei is negligible compared to their rest mass, their contribution to the pressure at zero temperature is insignificant. To determine the critical densities at which inverse $\beta$-decay becomes energetically favorable at the core of the WD, we numerically solve Eq.~(\ref{sec4:ineq2}) for stellar matter composed exclusively of carbon and oxygen ions.

\section{Modified TOV equations}\label{sec4}

The modified TOV equations in $f(R, T, L_m)= R+ \alpha TL_m$ gravity depend on the matter Lagrangian density $L_m$. For our WD study we will adopt the two choices of $L_m$ that the literature provides for an isotropic perfect fluid. In particular, for $L_m= p$, the stellar structure equations are given by \cite{Mota2024}
\begin{align}
    \frac{dm}{dr} &= 4\pi r^{2}\rho +\frac{\alpha r^2}{2}\left[ \frac{\rho}{2}(5p - \rho) + p^2 \right] ,  \label{TOVEq1p}  \\
    \frac{dp}{dr} &= -\frac{(\rho+ p) \left[4\pi r p +\frac{m}{r^2} + \frac{\alpha r}{4}(3p-\rho)p \right]}{\Big(1 - \frac{2m}{r}\Big) \left[1 + \frac{\alpha p}{16\pi + \alpha(5p - \rho)} \Big(1-\frac{d\rho}{dp}\Big) \right]} ,  \label{TOVEq2p}
\end{align}
while for $L_m= -\rho$ such equations take the form
\begin{align}
    \frac{dm}{dr} &= 4\pi r^2\rho + \frac{\alpha r^2}{4}(3p- \rho)\rho ,  \label{TOVEq1Mrho}  \\
    \frac{dp}{dr} &= -\frac{(\rho+ p) \left[4\pi r p +\frac{m}{r^2} + \frac{3\alpha r}{4}(p-\rho)p - \frac{\alpha r}{2}\rho^2 \right]}{\Big(1 - \frac{2m}{r}\Big)\left\{1 + \frac{\alpha\left[3p(1-d\rho/dp) - 4\rho (d\rho/dp) \right]}{16\pi + 3\alpha(p - \rho)} \right\}} ,  \label{TOVEq2Mrho}
\end{align}
where $m(r)$ stands for the gravitational mass within a sphere of radius $r$, and the relation $p= p(\rho)$ is the EoS that describes the microphysics of the WD. The new parameter $\alpha$ allows us to quantify the deviations of the different physical quantities with respect to their GR values (which are obtained when $\alpha \rightarrow 0$).

As usual in standard GR, both sets of differential equations are solved from the center at $r=0$ to the surface of the WD, where $r= r_{\rm sur}$. At the center ($r=0$) of the star, the mass and pressure satisfy the following boundary conditions:
\begin{align}\label{BoundCondEq}
    m(0) &= 0,  &  \rho(0) &= \rho_c , 
\end{align}
where $\rho_c$ is the central density and $r_{\rm sur}$ is determined when the pressure vanishes, i.e.~$p(r_{\rm sur})= 0$. Thus, we can determine the gravitational mass of the stars as $M= m(r_{\rm sur})$, and consequently sequences of WDs will be built by varying the central density given a specific $\alpha$ for the adopted gravity theory. These families of WDs will be represented in the well-known $M-r_{\rm sur}$ diagram.

\section{Mass-radius diagrams}\label{sec5}

\begin{figure*}
\includegraphics[width=17.4cm]{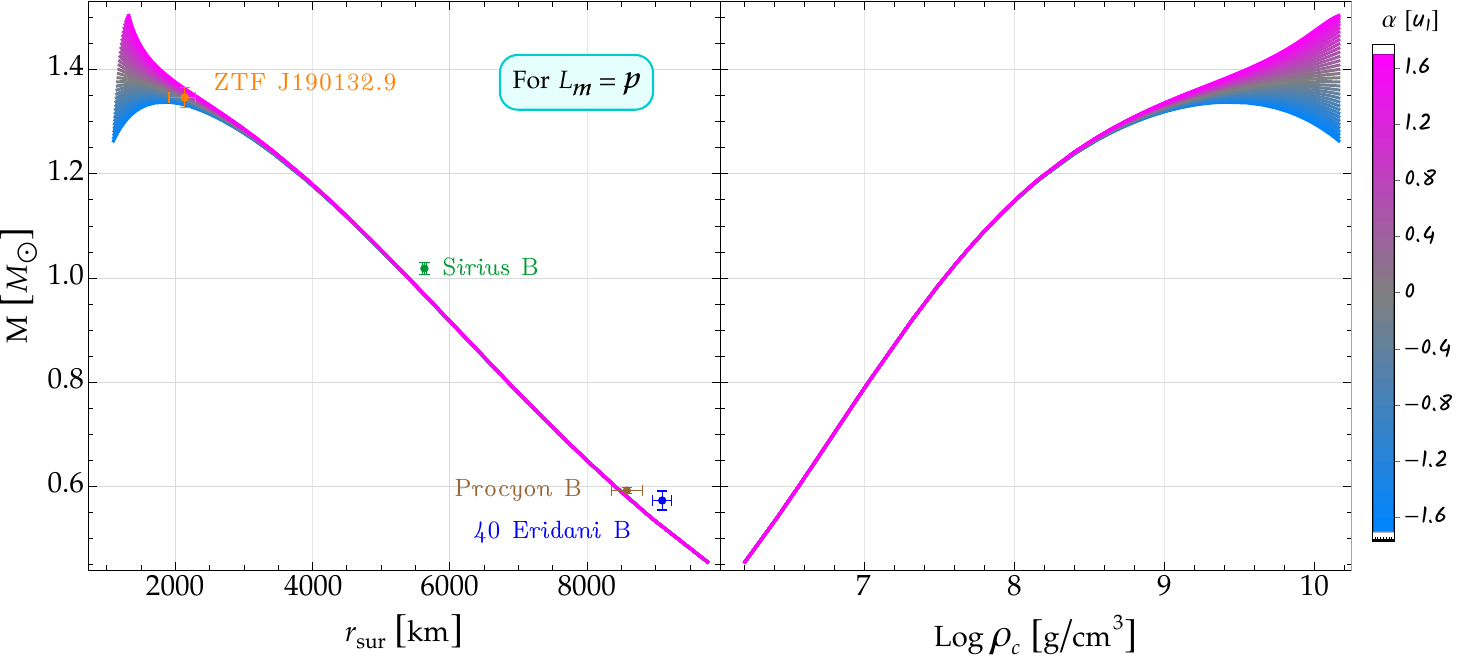}
\includegraphics[width=17.4cm]{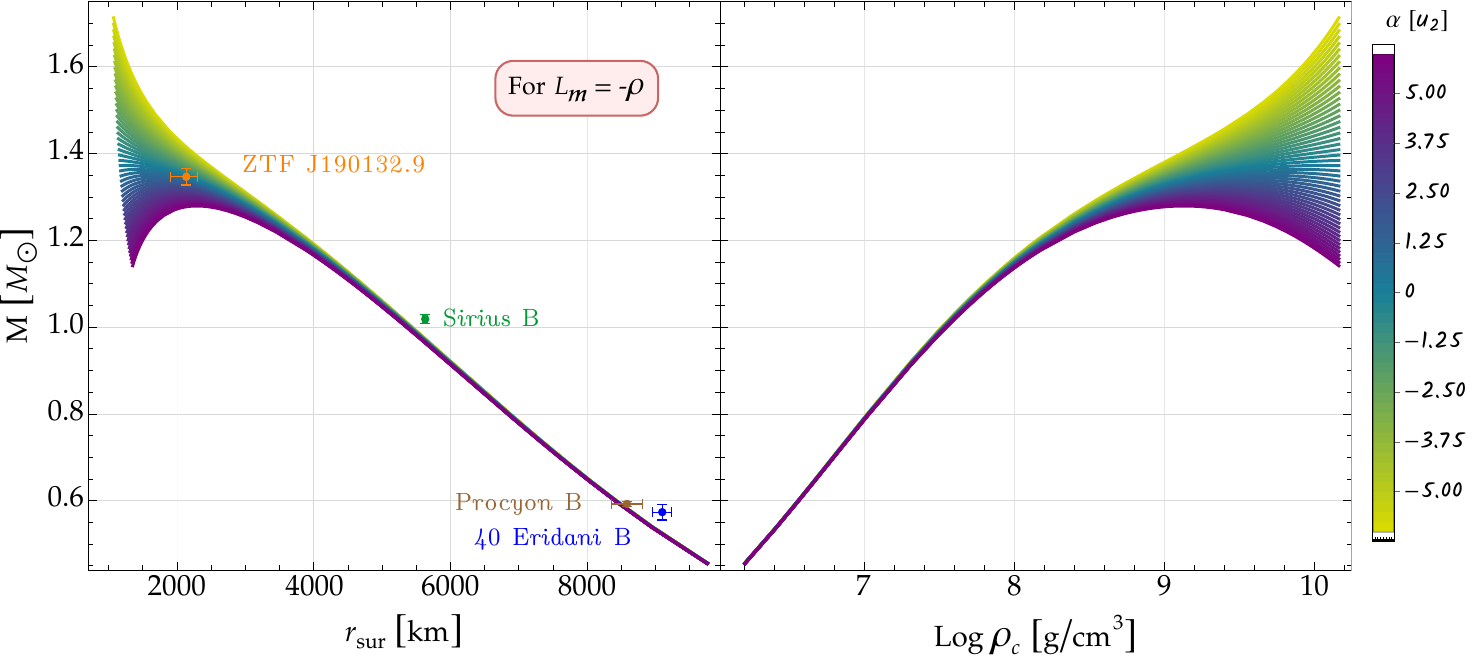}
    \caption{$M-r_{\rm sur}$ diagram (left) and $M-\rho_c$ relation (right) for WDs in $f(R, T, L_m)= R+ \alpha TL_m$ for several values of $\alpha$. The top panel is the result of solving the modified TOV equations \eqref{TOVEq1p} and \eqref{TOVEq2p}, where the coupling constant $\alpha$ has been varied in the range $\alpha \in [-1.7, 1.7]\, u_1$ with $u_1 = 10^{-73}\, \rm s^4/kg^2$. The bottom panel corresponds to the choice $L_m= -\rho$ where $\alpha \in [-6.0,6.0]\, u_2$ with $u_2$ being given by $u_2= 10^{-77}\, \rm s^4/kg^2$. We have included the observational mass–radius data from some well-studied and precisely constrained WDs: Sirius B, Procyon B, 40 Eridani B and ZTF J190132.9$+$145808.7 (see Table \ref{tabConstraints}). }%
    \label{FigMRCden}%
\end{figure*}

\subsection{Theoretical predictions}

To model the structure of WDs, we employ a realistic EoS that self-consistently incorporates both the pressure of a relativistic degenerate electron gas and the energy corrections associated with the ionic crystal lattice. After having chosen the matter Lagrangian density $L_m$ and specified the value of the coupling constant $\alpha$, we begin our analysis by constructing the $M-r_{\rm sur}$ relations by varying the central density $\rho_c$. Here, we vary \(\alpha\) only to map the phenomenology and identify the parameter domain compatible with massive WDs; observational data must ultimately select a single value (with uncertainties) for each adopted prescription of \(L_m\). Specifically, by solving the stellar structure equations \eqref{TOVEq1p} and \eqref{TOVEq2p} with initial conditions \eqref{BoundCondEq}, we obtain the upper plot in Fig.~\ref{FigMRCden} for $L_m= p$. We observe that the $\alpha TL_m$ term has a substantial effect on the massive WDs, i.e., at densities $\rho_c \gtrsim 10^9\, \rm g/cm^3$, while the impact of $\alpha$ is irrelevant at low central densities. In particular, a positive (negative) $\alpha$ increases (decreases) the gravitational mass of the WD relative to the general relativistic counterpart. For this choice of $L_m$, the parameter $\alpha$ has been given in $u_1 = 10^{-73}\, \rm s^4/kg^2$ units, namely, $10^5$ times larger than in the case of NSs and QSs as shown in our previous study \cite{Mota2024}. This means that in the case of WDs, larger values of $\alpha$ must be used to observe appreciable changes in the $M-r_{\rm sur}$ diagrams than in the case of NSs or QSs. This qualitative behavior is similar to that obtained in other modified gravity theories \cite{PRETEL2025}, such as in regularized 4D Einstein-Gauss-Bonnet gravity where the free parameter that quantifies the deviations from pure Einstein gravitation is larger in the case of WDs than in that of NSs. Again, here we attribute these differences on the order of $\alpha$ between WDs and NSs to the fact that we are dealing with different stellar systems, that is, with different energy density ranges.

\begin{figure*}
\includegraphics[width=8.4cm]{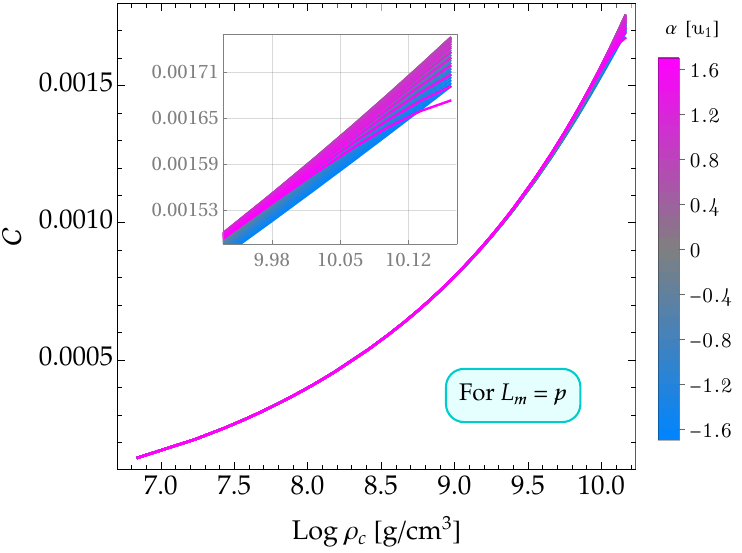}
\includegraphics[width=8.5cm]{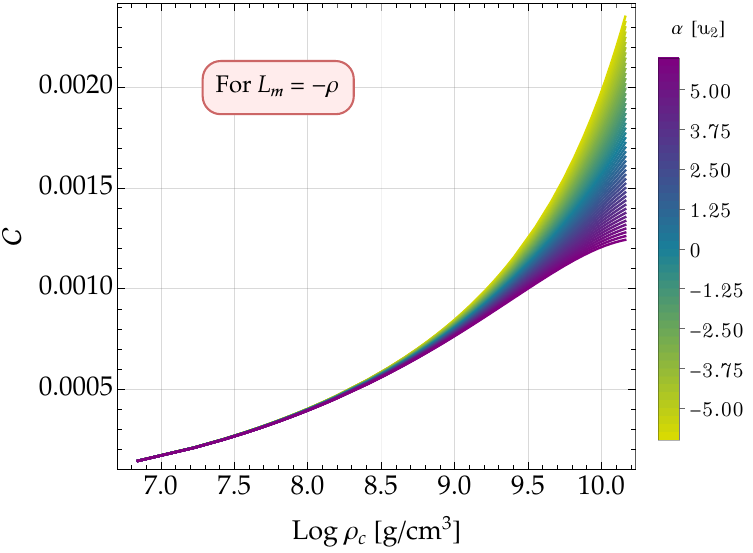}
    \caption{Compactness versus central density relation for the WD configurations shown in Fig.~\ref{FigMRCden}. One observes that the two choices of Lagrangian density lead to remarkably different compactnesses. }%
    \label{FigComp}%
\end{figure*}

According to the top-left plot of Fig.~\ref{FigMRCden}, for high central densities, the radius of the star increases as $\alpha$ increases from its negative values. As a consequence, this generates a peculiar behavior in the compactness, defined as $\mathcal{C}= M/r_{\rm sur}$, when it is plotted as a function of the central density in the left panel of Fig.~\ref{FigComp}. Indeed, the compactness increases to a certain maximum value and then begins to decrease for some positive values of $\alpha$. Nevertheless, changes in $\mathcal{C}$ due to the modified gravity term $\alpha TL_m$ are irrelevant in the low central density branch when $L_m= p$. 

In a similar way we numerically solve the modified TOV equations for $L_m= -\rho$, i.e., the differential equations \eqref{TOVEq1Mrho} and \eqref{TOVEq2Mrho}. Our results for this case are shown in the lower panel of Fig.~\ref{FigMRCden}, where we have considered the range $\vert\alpha\vert \leq 6.0\, u_2$ with $u_2$ given by $u_2= 10^{-77}\, \rm s^4/kg^2$, indicating that now our $\alpha$ is 100 times larger than in the context of NSs and QSs \cite{Mota2024}. Here, the largest changes take place at central densities $\rho_c \gtrsim 10^8\, \rm g/cm^3$, where positive (negative) values of $\alpha$ decrease (increase) the maximum mass, and which is opposite to the effect generated by the choice $L_m= p$. Likewise, the compactness in the right plot of Fig.~\ref{FigComp} exhibits a different behavior from that produced by $L_m= p$. For the choice $L_m= -\rho$, $\mathcal{C}$ always increases as $\alpha$ decreases for high central densities. Specifically, the maximum compactness obtained in GR $\mathcal{C}_{\rm max} \approx 0.00175$ can increase to $\mathcal{C}_{\rm max} \approx 0.00235$ when $\alpha= 6.0\, u_2$.

A remarkable result of this work is that both choices of $L_m$ lead to a significant modification of the usual Chandrasekhar mass limit, thus favoring the description of massive WDs. Even more interesting is the fact that for some values of $\alpha$ it is not possible to find a critical WD, that is, a star of maximum mass indicating the transition from stability to instability according to the usual criterion for stability $dM/d\rho_c > 0$. For example, for $L_m=p$, negative values of $\alpha$ allow us to find a critical configuration such that $dM/d\rho_c = 0$ on the $M(\rho_c)$-curves, while for $\alpha$ sufficiently large and positive such a critical WD is not found. For $L_m= -\rho$, this behavior is opposite. In summary, depending on the value of $\alpha$ for each choice, it is possible to obtain stable super-Chandrasekhar WDs within the context of $f(R, T, L_m)= R+ \alpha TL_m$ gravity; a result not expected in WDs described by pure GR.

Our results are in qualitative agreement with previous studies of WDs in regularized 4D Einstein-Gauss-Bonnet gravity \cite{PRETEL2025} and in Palatini $f(R)$ gravity \cite{Sarmah2022PRD, Kalita2023}. In all these gravitational frameworks, deviations from GR arise predominantly in the high-density regions of relativistic WDs, while remaining negligible in the low-mass regime. Furthermore, depending on the values of the free parameters of the underlying gravity theory, these models—including the one considered here—admit massive stellar configurations without the appearance of a maximum mass beyond which stability is lost. Nevertheless, despite these similarities across different modified gravity theories, our results differ markedly from those obtained in $f(R,T)=R+2\beta T$ gravity \cite{Carvalho:2017pgk}, where the additional term $2\beta T$ induces the largest deviations at large stellar radii, corresponding to low central densities.

\subsection{Bayesian constraints on $\alpha$ from WD mass-radius observations}

In this work we perform Bayesian inference to test and constrain the theoretical model and to optimize the parameter $\alpha$. Given a model $\mathcal{M}$, in our case two models: $L_m=p$ and $L_m=-\rho$, described by a set of parameters $\boldsymbol{\theta}=\alpha$, Bayesian inference provides a probabilistic description of the region of parameter space consistent with the data. Prior knowledge about the parameters is encoded in the prior distribution $\mathcal{P}(\boldsymbol{\theta}, \mathcal{M})$, which represents our state of information before considering the observations. The information provided by the observational constraints $\mathcal{D}$ enters through the likelihood function $\mathcal{L}(\mathfrak{D}|\boldsymbol{\theta},\mathcal{M})$, which quantifies the probability of obtaining the observed data for a given choice of parameters. Bayes’ theorem combines the prior and the likelihood to yield the posterior probability distribution $\mathcal{P}(\boldsymbol{\theta}|\mathfrak{D},\mathcal{M}) \propto \mathcal{L}(\mathfrak{D}|\boldsymbol{\theta},\mathcal{M}) \mathcal{P}(\boldsymbol{\theta}, \mathcal{M})$, which represents the probability density function of the parameters conditioned on the data~\cite{sivia2006data}. In our case, the posterior distribution is obtained using a Markov chain Monte Carlo (MCMC) approach, specifically the Goodman and Weare's Affine Invariant MCMC~\cite{goodman2010ensemble}, as implemented in the \texttt{emcee} package~\cite{foreman2013emcee}.

In order to obtain the most accurate estimate of the parameter $\alpha$ for each matter Lagrangian density, we performed our Bayesian inference using a Gaussian likelihood function. For such inference, we consider observational $M-r_{\rm sur}$ data from three well studied and precisely constrained WDs: Sirius B, 40 Eridani B, and Procyon B. Additionally, we also included the massive white dwarf ZTF J190132.9$+$145808.7, with a mass close to $1.4\, M_{\odot}$, where our gravity model exhibits more pronounced deviations from GR. The masses and radii, along with their respective uncertainties, for all the WDs considered in this study are summarized in Table \ref{tabConstraints}. The observational sample used in this analysis was intentionally restricted to a small set of benchmark WDs with independently determined and highly precise mass-radius measurements. Rather than conducting a population-level study, our goal is to probe the phenomenological impact of the matter-geometry coupling parameter using objects that span distinct regions of the mass-radius plane while minimizing systematic uncertainties associated with heterogeneous catalogues. In this context, Sirius~B, Procyon~B, and 40~Eridani~B serve as well-calibrated anchors in the low- and intermediate-mass regimes, whereas ZTF~J190132.9+145808.7 probes the ultra-massive domain where deviations from conventional GR are expected to be most pronounced. A comprehensive statistical analysis using larger samples will be considered in future work.

\begin{table}[t]
    \centering
        \caption{Observational mass–radius measurements of WDs employed as constraints. }
    \begin{ruledtabular}
    \begin{tabular}{l c c}
         \textbf{Star}  &  \textbf{Mass}  &  \textbf{Radius}  \\
         \hline
         Sirius B \cite{bond2017sirius}  &  $1.018 \pm 0.011\, M_{\odot}$  &  $5634 \pm 32$ km\\

         40 Eridani B \cite{bond2017astrophysical}  &   $0.573 \pm 0.018\, M_{\odot}$  &  $9100 \pm 139$ km  \\

         Procyon B \cite{bond2015hubble,provencal2002procyon}  &  $	0.592 \pm 0.006\, M_{\odot}$  &  $8585 \pm 223$ km  \\ 

         ZTF J190132.9 \cite{caiazzo2021highly}  &  $1.346 \pm 0.019\, M_{\odot}$  &  $2140_{-230}^{+160}$ km 
    \end{tabular}
    \end{ruledtabular}
    \label{tabConstraints}
\end{table}

\begin{figure}
    \centering
    \includegraphics[width=8.5cm]{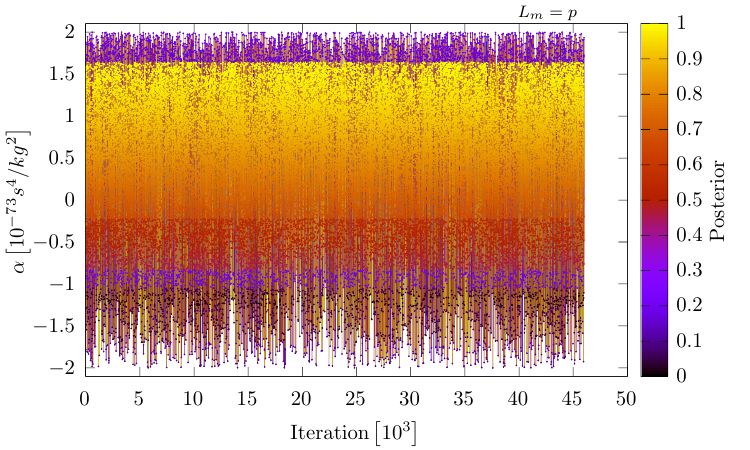}\\
    \includegraphics[width=8.5cm]{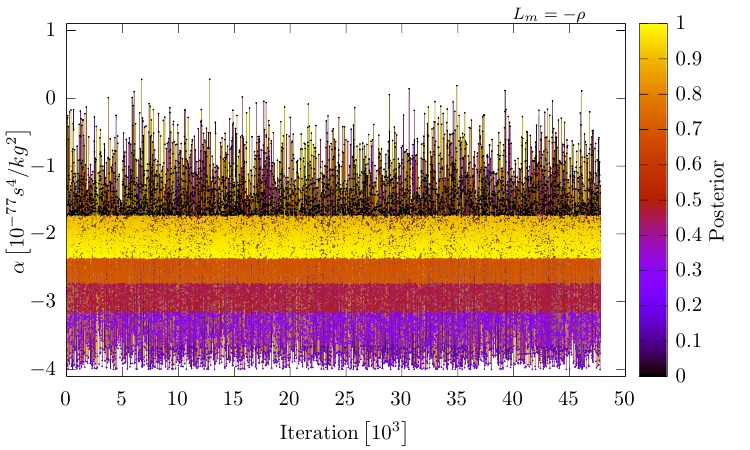}
    \caption{Trace plot of the MCMC samples for the coupling parameter $\alpha$ using both choices; $L_m=p$ (top panel) and $L_m=-\rho$ (bottom panel). }
    \label{figtrace}
\end{figure}

In our inferences, we assumed uniform priors. For the model with $L_m = p$, the parameter $\alpha$ was allowed to vary within the range 
\begin{equation}
    \alpha \in \left[-2,2\right] \times 10^{-73} \, \mathrm{s^4\,kg^{-2}},
\end{equation}
while for the model with $L_m = -\rho$, the range was set to 
\begin{equation}
    \alpha \in \left[-4,1\right] \times 10^{-77} \, \mathrm{s^4\,kg^{-2}}.
\end{equation}
The domain of our prior ranges can be justified by analyzing the trace plots of each inference in Fig.~\ref{figtrace}, where we can observe that the posterior samples remain well within the prior bounds and do not accumulate near the boundaries, indicating that the inferences are not driven by prior truncation. From Fig.~\ref{figtrace} we can also conclude that the trace plot of $\alpha$ shows stable fluctuations around a constant mean with no visible trends, implying good mixing and convergence of the MCMC chain.

\begin{figure}
    \centering
    \includegraphics[scale=0.3]{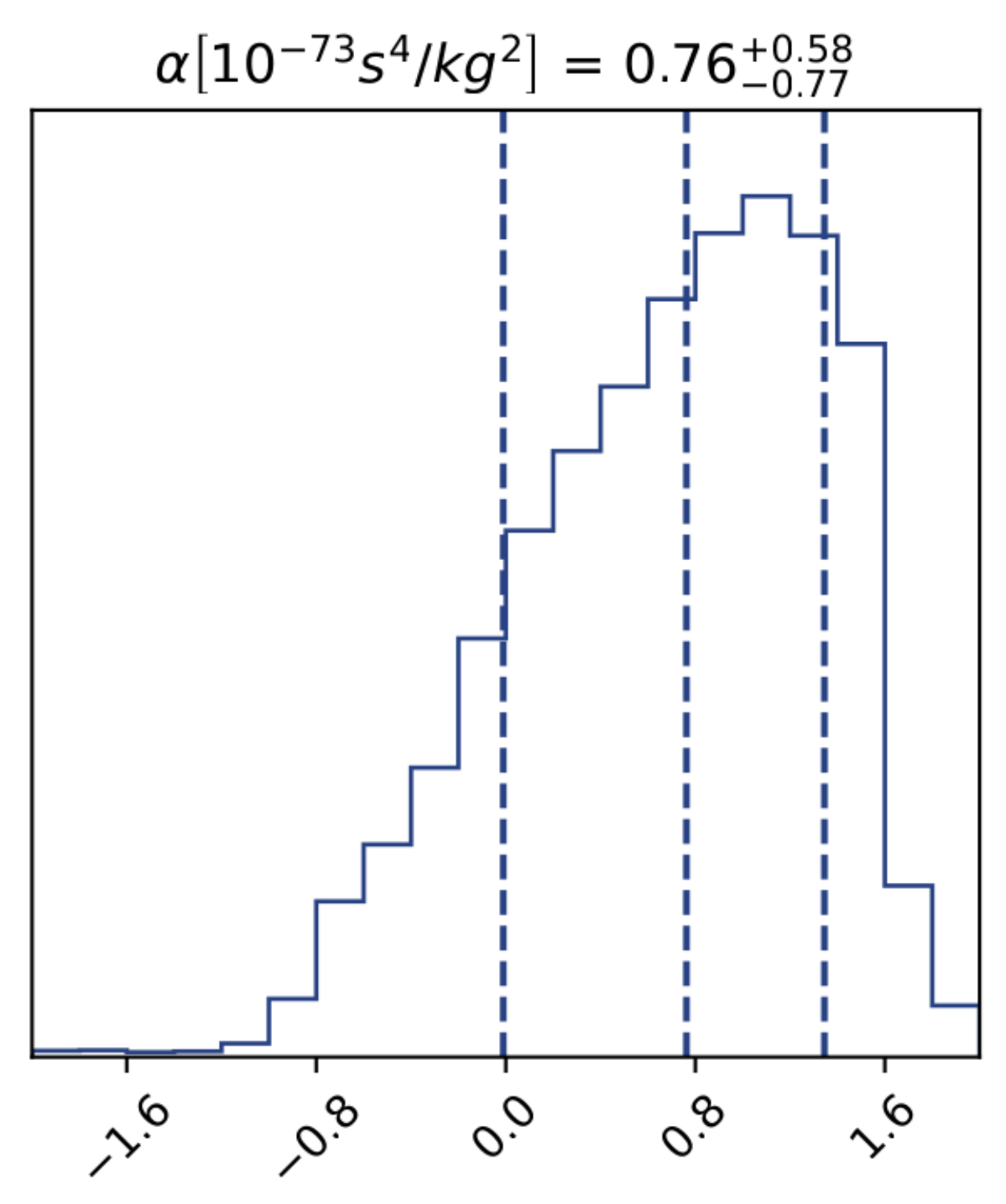}\\
    \includegraphics[scale=0.3]{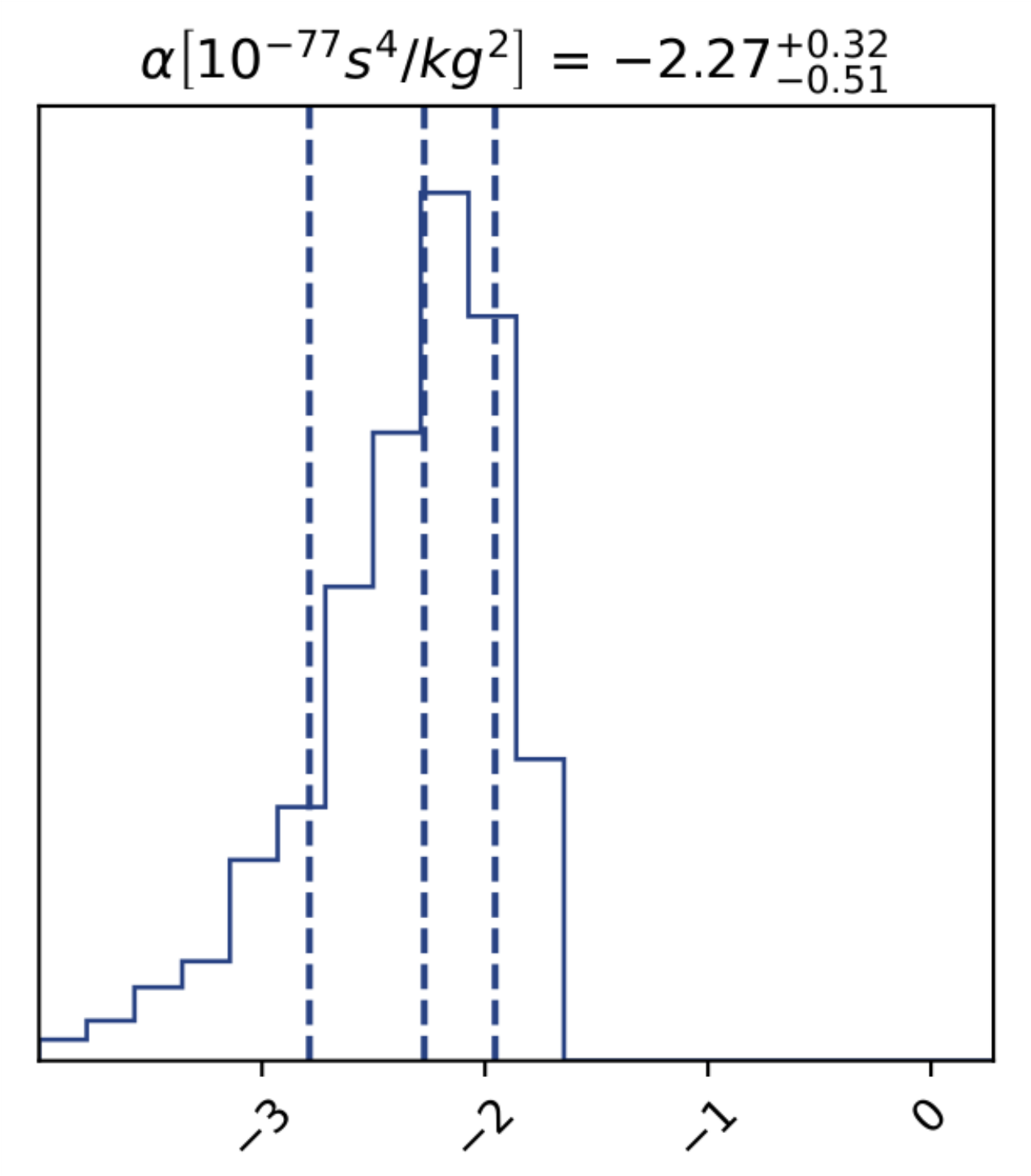}
    \caption{Histograms showing the posterior distributions of the parameter $\alpha$ for the $L_m=p$ model on the top and the $L_m=-\rho$ model on the bottom. The dashed vertical lines represent the 0.16, 0.5, and 0.84 quantiles.}
    \label{fighist}
\end{figure}

Based on our inference, Fig.~\ref{fighist}, the most probable values of $\alpha$, given the observational constraints and the EoS adopted in this work, are
\begin{equation}
    \alpha = \begin{cases} \left( 0.76^{+0.58}_{-0.77} \right) \times 10^{-73} \, \mathrm{s^4\,kg^{-2}}  &  \text{for } \ L_m = p , \\
    \left(-2.27^{+0.32}_{-0.51}\right) \times 10^{-77} \, \mathrm{s^4\,kg^{-2}} &  \text{for } \ L_m = -\rho . \end{cases}
\end{equation}

\begin{figure}
    \centering
    \includegraphics[width=8.5cm]{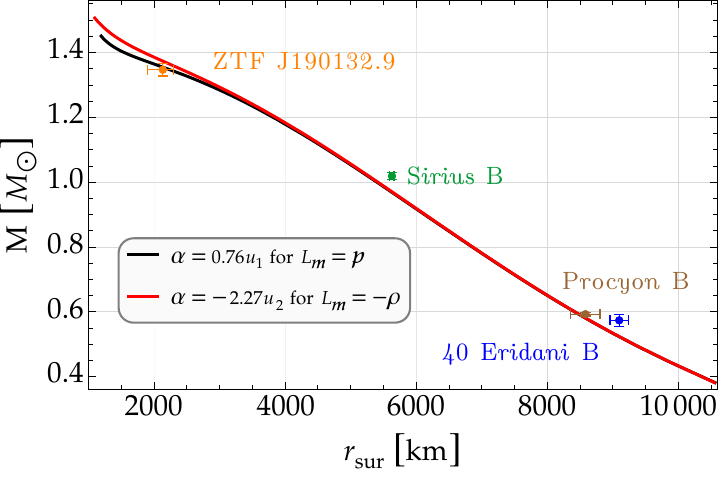}
    \caption{WD mass-radius relations in $f(R,T,L_m)=R+\alpha T L_m$ gravity for the values of $\alpha$ favored by the Bayesian analysis of the observational data. }
    \label{fig:placeholder}
\end{figure}

Figure \ref{fig:placeholder} shows the WD mass-radius relation obtained using the values of the coupling parameter~$\alpha$ inferred from the observational data. The theoretical $M-r_{\rm sur}$ curves correspond to the posterior-preferred values of~$\alpha$ for each choice of the matter Lagrangian density, while the observational points indicate the measured masses and radii with their associated uncertainties. For both prescriptions, $L_m = p$ and $L_m = -\rho$, the inferred $M-r_{\rm sur}$ curves are consistent with the observational constraints within the credible intervals. At low and intermediate masses, the predicted radii remain close to the GR expectation, indicating that the matter-geometry coupling does not introduce significant deviations in this regime. This confirms that the inferred values of~$\alpha$ preserve the well-tested structure of canonical WDs. In contrast, deviations from GR become progressively more relevant at higher masses, where the $M-r_{\rm sur}$ relation exhibits a systematic shift that improves agreement with massive and ultra-massive WDs. In particular, the inclusion of ZTF~J190132.9$+$145808.7, with a mass close to the Chandrasekhar limit, significantly constrains the posterior distribution of~$\alpha$, anchoring the theoretical curve in the regime where modified-gravity effects are expected to be maximal.

From a statistical standpoint, the observational points are not preferentially aligned with a star-dependent value of~$\alpha$, but instead cluster around a single mass-radius sequence for each choice of~$L_m$. This supports the interpretation of~$\alpha$ as a fundamental parameter of the gravity theory rather than an effective, object-dependent quantity. The posterior uncertainties translate into a narrow band in the $M-r_{\rm sur}$ plane, indicating that the observational data are sufficiently informative to constrain the coupling parameter at WD densities.

Overall, our findings demonstrate that a single, statistically inferred value of~$\alpha$ is able to reproduce simultaneously the properties of both standard and massive WDs, while preserving consistency with standard GR in the low-density regime. This highlights the mass-radius relation as a sensitive probe of matter-geometry couplings and provides observational support for the viability of $f(R,T,L_m)$ gravity at the scale of WD interiors.

\section{Conclusions and future perspectives}\label{sec6}

The purpose of this work has been to examine the relativistic structure and stability of WDs in $f(R,T,L_m)$ gravity, assuming a realistic EoS for the microphysics of such stars. In particular, the $f(R, T, L_m)= R+ \alpha TL_m$ gravity model has been employed to address the macrophysics of WDs, where $\alpha$ is a matter-geometry coupling and measures the deviations from the usual GR. We have therefore focused on studying the effect of such a parameter on the most basic global properties of a compact star: its radius and mass. Our findings reveal that the canonical Chandrasekhar mass limit can be substantially modified due to the presence of the $\alpha TL_m$ term, which would strongly favor the observational evidence of super-Chandrasekhar WDs.

For comparison purposes, our work has adopted both choices for the matter Lagrangian density. Specifically, for $L_m= p$, the WD mass increases as a consequence of increasing the value of $\alpha$ from its negative values, mainly in the high-central-density branch where $\rho_c \gtrsim 10^9\, \rm g/cm^3$. Nonetheless, for $L_m= -\rho$, the impact of the coupling constant is opposite; the gravitational mass increases (decreases) for negative (positive) $\alpha$. Remarkably, according to the classical criterion for stellar stability $dM/d\rho_c > 0$ where the maximum mass corresponds to a transition point from stability to instability, it is not possible to find a maximum for some values of $\alpha$, suggesting that for example the WDs belonging to the magenta curves (for $L_m= p$) and yellow curves (for $L_m= -\rho$) are always stable. In other words, our study shows that super-Chandrasekhar WDs can be consistently described as stable massive WDs in $f(R,T,L_m)$ gravity.

In summary, the Bayesian analysis showed that WD mass-radius observations consistently favor a single value of the coupling parameter $\alpha$ for each choice of $L_m$. This supports its interpretation as a fundamental parameter of the gravitational theory and demonstrates that $f(R,T,L_m)$ gravity can simultaneously account for both standard and massive WDs while remaining compatible with conventional GR at low densities.

The EoS adopted in this work, a relativistic degenerate electron gas augmented by body-centred-cubic (bcc) lattice Coulomb corrections marks a notable advance over the classical polytropic prescriptions still common in Chandrasekhar-style analyses (\(\gamma = 5/3\) in the non-relativistic limit and \(\gamma = 4/3\) in the ultra-relativistic regime). Explicit inclusion of the lattice pressure term $p_{L} \;\propto\; n_e^{4/3}\, Z^{2/3}$, captures the EoS softening at intermediate densities and therefore yields more realistic mass-radius relations for carbon-rich white dwarfs.

Two idealizations nevertheless remain: 
\begin{enumerate}
  \item \textit{Zero-temperature assumption} (\(T = 0\,\mathrm{K}\)). Finite-\(T\) effects become important in the outer envelopes and during crystallisation. Thermodynamic tables such as can be seen in \cite{1999A&A...346..345P}, which include \(e\!-\!e\) and \(e\!-\!\text{ion}\) interactions as well as explicit \(T\)-dependence, have already been employed to bracket the allowable mass range of rotating WDs; their adoption would enable a systematic analysis of how cooling modifies the mass–radius curve and the onset of inverse-\(\beta\) instabilities.
  \item \textit{Neglect of magnetic fields}. Fields stronger than \(B \gtrsim 10^{13}\,\mathrm{G}\) quantize Landau levels, introduce pressure anisotropy, and can raise the Chandrasekhar limit. GR–Maxwell studies with bcc lattices indicate \(M_{\max} \simeq 2\text{--}2.2\,M_\odot\) for poloidal configurations [see \cite{Otoniel_2019}]. Implementing a self-consistent magnetic field in the modified TOV framework would clarify the interplay between matter–curvature coupling (\(\alpha\)) and magnetohydrostatic support.
\end{enumerate}

Looking ahead, several avenues merit exploration: 
\begin{itemize}
  \item \emph{Multi-component thermal EoSs (He--C--O).} Examine how chemical stratification affects the minimum radius and the \(\beta\)-decay threshold using mixed liquid/crystalline plasma models.

  \item \emph{Strong magnetic fields.} Solve the coupled TOV--Maxwell equations in \(f(R,T,L_{m})\) gravity, including Landau quantization, anisotropic pressure and stellar deformations (prolate/oblate), following the methodology of Otoniel \emph{et al.} \cite{Otoniel_2019}.

  \item \emph{Additional microphysics.} Incorporate inverse-\(\beta\) reactions and pycnonuclear fusion at \(B>0\) and \(T>0\) to delimit dynamical stability limits.
\end{itemize}

Altogether, replacing the present zero-\(T\), field-free EoS with a thermally and magnetically enriched description will provide a stringent test of the robustness of the \(\alpha TL_{m}\) coupling against the micro and macrophysical processes shaping extreme WDs.

\begin{acknowledgments}
JMZP acknowledges support from ``Fundação Carlos Chagas Filho de Amparo à Pesquisa do Estado do Rio de Janeiro'' -- FAPERJ, Process SEI-260003/000308/2024. E. Otoniel acknowledges support from FUNCAP(BP6-0241-00335.01.00/25). C. Flores  acknowledges the financial support of the productivity program of the conselho Nacional de Desenvolvimento Científico e Tecnológico (CNPq), with Project No. 304569/2022-4. 
F.M.S. would like to thank CNPq for financial support under research project No. 403007/2024-0 and research fellowship No. 201145/2025-1.
\end{acknowledgments}


\newpage
\begin{thebibliography}{79}%
\makeatletter
\providecommand \@ifxundefined [1]{%
 \@ifx{#1\undefined}
}%
\providecommand \@ifnum [1]{%
 \ifnum #1\expandafter \@firstoftwo
 \else \expandafter \@secondoftwo
 \fi
}%
\providecommand \@ifx [1]{%
 \ifx #1\expandafter \@firstoftwo
 \else \expandafter \@secondoftwo
 \fi
}%
\providecommand \natexlab [1]{#1}%
\providecommand \enquote  [1]{``#1''}%
\providecommand \bibnamefont  [1]{#1}%
\providecommand \bibfnamefont [1]{#1}%
\providecommand \citenamefont [1]{#1}%
\providecommand \href@noop [0]{\@secondoftwo}%
\providecommand \href [0]{\begingroup \@sanitize@url \@href}%
\providecommand \@href[1]{\@@startlink{#1}\@@href}%
\providecommand \@@href[1]{\endgroup#1\@@endlink}%
\providecommand \@sanitize@url [0]{\catcode `\\12\catcode `\$12\catcode `\&12\catcode `\#12\catcode `\^12\catcode `\_12\catcode `\%12\relax}%
\providecommand \@@startlink[1]{}%
\providecommand \@@endlink[0]{}%
\providecommand \url  [0]{\begingroup\@sanitize@url \@url }%
\providecommand \@url [1]{\endgroup\@href {#1}{\urlprefix }}%
\providecommand \urlprefix  [0]{URL }%
\providecommand \Eprint [0]{\href }%
\providecommand \doibase [0]{http://dx.doi.org/}%
\providecommand \selectlanguage [0]{\@gobble}%
\providecommand \bibinfo  [0]{\@secondoftwo}%
\providecommand \bibfield  [0]{\@secondoftwo}%
\providecommand \translation [1]{[#1]}%
\providecommand \BibitemOpen [0]{}%
\providecommand \bibitemStop [0]{}%
\providecommand \bibitemNoStop [0]{.\EOS\space}%
\providecommand \EOS [0]{\spacefactor3000\relax}%
\providecommand \BibitemShut  [1]{\csname bibitem#1\endcsname}%
\let\auto@bib@innerbib\@empty
\bibitem [{\citenamefont {{Einstein}}(1915)}]{1915SPAW831E}%
  \BibitemOpen
  \bibfield  {author} {\bibinfo {author} {\bibfnamefont {A.}~\bibnamefont {{Einstein}}},\ }\href@noop {} {\bibfield  {journal} {\bibinfo  {journal} {Sitzungsberichte der K{\"o}niglich Preu{\ss}ischen Akademie der Wissenschaften (Berlin)}\ ,\ \bibinfo {pages} {831}} (\bibinfo {year} {1915})}\BibitemShut {NoStop}%
\bibitem [{\citenamefont {Abbott}\ \emph {et~al.}(2016)\citenamefont {Abbott} \emph {et~al.}}]{GWblackhole}%
  \BibitemOpen
  \bibfield  {author} {\bibinfo {author} {\bibfnamefont {B.~P.}\ \bibnamefont {Abbott}} \emph {et~al.} (\bibinfo {collaboration} {LIGO Scientific, Virgo}),\ }\href {\doibase 10.1103/PhysRevLett.116.061102} {\bibfield  {journal} {\bibinfo  {journal} {Phys. Rev. Lett.}\ }\textbf {\bibinfo {volume} {116}},\ \bibinfo {pages} {061102} (\bibinfo {year} {2016})}\BibitemShut {NoStop}%
\bibitem [{\citenamefont {Abbott}\ and\ \citenamefont {et~al.}(2017)}]{PhysRevLett.119.161101}%
  \BibitemOpen
  \bibfield  {author} {\bibinfo {author} {\bibfnamefont {B.~P.}\ \bibnamefont {Abbott}}\ and\ \bibinfo {author} {\bibnamefont {et~al.}} (\bibinfo {collaboration} {LIGO Scientific Collaboration and Virgo Collaboration}),\ }\href {\doibase 10.1103/PhysRevLett.119.161101} {\bibfield  {journal} {\bibinfo  {journal} {Phys. Rev. Lett.}\ }\textbf {\bibinfo {volume} {119}},\ \bibinfo {pages} {161101} (\bibinfo {year} {2017})}\BibitemShut {NoStop}%
\bibitem [{\citenamefont {{Event Horizon Telescope Collaboration}}(2019)}]{2019ApJ...875L...1E}%
  \BibitemOpen
  \bibfield  {author} {\bibinfo {author} {\bibnamefont {{Event Horizon Telescope Collaboration}}},\ }\href {\doibase 10.3847/2041-8213/ab0ec7} {\bibfield  {journal} {\bibinfo  {journal} {ApJL}\ }\textbf {\bibinfo {volume} {875}},\ \bibinfo {eid} {L1} (\bibinfo {year} {2019})}\BibitemShut {NoStop}%
\bibitem [{\citenamefont {Nojiri}\ and\ \citenamefont {Odintsov}(2003)}]{PRD03sergei}%
  \BibitemOpen
  \bibfield  {author} {\bibinfo {author} {\bibfnamefont {S.}~\bibnamefont {Nojiri}}\ and\ \bibinfo {author} {\bibfnamefont {S.~D.}\ \bibnamefont {Odintsov}},\ }\href {\doibase 10.1103/PhysRevD.68.123512} {\bibfield  {journal} {\bibinfo  {journal} {Phys. Rev. D}\ }\textbf {\bibinfo {volume} {68}},\ \bibinfo {pages} {123512} (\bibinfo {year} {2003})}\BibitemShut {NoStop}%
\bibitem [{\citenamefont {Allemandi}\ \emph {et~al.}(2005)\citenamefont {Allemandi}, \citenamefont {Borowiec}, \citenamefont {Francaviglia},\ and\ \citenamefont {Odintsov}}]{Allemandi}%
  \BibitemOpen
  \bibfield  {author} {\bibinfo {author} {\bibfnamefont {G.}~\bibnamefont {Allemandi}}, \bibinfo {author} {\bibfnamefont {A.}~\bibnamefont {Borowiec}}, \bibinfo {author} {\bibfnamefont {M.}~\bibnamefont {Francaviglia}}, \ and\ \bibinfo {author} {\bibfnamefont {S.~D.}\ \bibnamefont {Odintsov}},\ }\href {\doibase 10.1103/PhysRevD.72.063505} {\bibfield  {journal} {\bibinfo  {journal} {Phys. Rev. D}\ }\textbf {\bibinfo {volume} {72}},\ \bibinfo {pages} {063505} (\bibinfo {year} {2005})}\BibitemShut {NoStop}%
\bibitem [{\citenamefont {Koyama}(2016)}]{Koyama2016}%
  \BibitemOpen
  \bibfield  {author} {\bibinfo {author} {\bibfnamefont {K.}~\bibnamefont {Koyama}},\ }\href {\doibase 10.1088/0034-4885/79/4/046902} {\bibfield  {journal} {\bibinfo  {journal} {Rep. Prog. Phys.}\ }\textbf {\bibinfo {volume} {79}},\ \bibinfo {pages} {046902} (\bibinfo {year} {2016})}\BibitemShut {NoStop}%
\bibitem [{\citenamefont {Nojiri}\ \emph {et~al.}(2017)\citenamefont {Nojiri}, \citenamefont {Odintsov},\ and\ \citenamefont {Oikonomou}}]{repsergei}%
  \BibitemOpen
  \bibfield  {author} {\bibinfo {author} {\bibfnamefont {S.}~\bibnamefont {Nojiri}}, \bibinfo {author} {\bibfnamefont {S.~D.}\ \bibnamefont {Odintsov}}, \ and\ \bibinfo {author} {\bibfnamefont {V.~K.}\ \bibnamefont {Oikonomou}},\ }\href {\doibase 10.1016/j.physrep.2017.06.001} {\bibfield  {journal} {\bibinfo  {journal} {Phys. Rept.}\ }\textbf {\bibinfo {volume} {692}},\ \bibinfo {pages} {1} (\bibinfo {year} {2017})}\BibitemShut {NoStop}%
\bibitem [{\citenamefont {Shankaranarayanan}\ and\ \citenamefont {Johnson}(2022)}]{Shankaranarayanan2022}%
  \BibitemOpen
  \bibfield  {author} {\bibinfo {author} {\bibfnamefont {S.}~\bibnamefont {Shankaranarayanan}}\ and\ \bibinfo {author} {\bibfnamefont {J.~P.}\ \bibnamefont {Johnson}},\ }\href {\doibase 10.1007/s10714-022-02927-2} {\bibfield  {journal} {\bibinfo  {journal} {Gen. Relativ. Gravit.}\ }\textbf {\bibinfo {volume} {54}},\ \bibinfo {pages} {44} (\bibinfo {year} {2022})}\BibitemShut {NoStop}%
\bibitem [{\citenamefont {de~Rham}\ \emph {et~al.}(2011)\citenamefont {de~Rham}, \citenamefont {Gabadadze},\ and\ \citenamefont {Tolley}}]{deRham:2010kj}%
  \BibitemOpen
  \bibfield  {author} {\bibinfo {author} {\bibfnamefont {C.}~\bibnamefont {de~Rham}}, \bibinfo {author} {\bibfnamefont {G.}~\bibnamefont {Gabadadze}}, \ and\ \bibinfo {author} {\bibfnamefont {A.~J.}\ \bibnamefont {Tolley}},\ }\href {\doibase 10.1103/PhysRevLett.106.231101} {\bibfield  {journal} {\bibinfo  {journal} {Phys. Rev. Lett.}\ }\textbf {\bibinfo {volume} {106}},\ \bibinfo {pages} {231101} (\bibinfo {year} {2011})}\BibitemShut {NoStop}%
\bibitem [{\citenamefont {Brans}\ and\ \citenamefont {Dicke}(1961)}]{Brans:1961sx}%
  \BibitemOpen
  \bibfield  {author} {\bibinfo {author} {\bibfnamefont {C.}~\bibnamefont {Brans}}\ and\ \bibinfo {author} {\bibfnamefont {R.~H.}\ \bibnamefont {Dicke}},\ }\href {\doibase 10.1103/PhysRev.124.925} {\bibfield  {journal} {\bibinfo  {journal} {Phys. Rev.}\ }\textbf {\bibinfo {volume} {124}},\ \bibinfo {pages} {925} (\bibinfo {year} {1961})}\BibitemShut {NoStop}%
\bibitem [{\citenamefont {Buchdahl}(1970)}]{1970MNRAS.150....1B}%
  \BibitemOpen
  \bibfield  {author} {\bibinfo {author} {\bibfnamefont {H.~A.}\ \bibnamefont {Buchdahl}},\ }\href {\doibase 10.1093/mnras/150.1.1} {\bibfield  {journal} {\bibinfo  {journal} {MNRAS}\ }\textbf {\bibinfo {volume} {150}},\ \bibinfo {pages} {1} (\bibinfo {year} {1970})}\BibitemShut {NoStop}%
\bibitem [{\citenamefont {De~Felice}\ and\ \citenamefont {Tsujikawa}(2010)}]{DeFelice:2010aj}%
  \BibitemOpen
  \bibfield  {author} {\bibinfo {author} {\bibfnamefont {A.}~\bibnamefont {De~Felice}}\ and\ \bibinfo {author} {\bibfnamefont {S.}~\bibnamefont {Tsujikawa}},\ }\href {\doibase 10.12942/lrr-2010-3} {\bibfield  {journal} {\bibinfo  {journal} {Living Rev. Rel.}\ }\textbf {\bibinfo {volume} {13}},\ \bibinfo {pages} {3} (\bibinfo {year} {2010})}\BibitemShut {NoStop}%
\bibitem [{\citenamefont {Capozziello}\ and\ \citenamefont {De~Laurentis}(2011)}]{Capozziello:2011et}%
  \BibitemOpen
  \bibfield  {author} {\bibinfo {author} {\bibfnamefont {S.}~\bibnamefont {Capozziello}}\ and\ \bibinfo {author} {\bibfnamefont {M.}~\bibnamefont {De~Laurentis}},\ }\href {\doibase 10.1016/j.physrep.2011.09.003} {\bibfield  {journal} {\bibinfo  {journal} {Phys. Rept.}\ }\textbf {\bibinfo {volume} {509}},\ \bibinfo {pages} {167} (\bibinfo {year} {2011})}\BibitemShut {NoStop}%
\bibitem [{\citenamefont {Nojiri}\ and\ \citenamefont {Odintsov}(2011)}]{Nojiri:2010wj}%
  \BibitemOpen
  \bibfield  {author} {\bibinfo {author} {\bibfnamefont {S.}~\bibnamefont {Nojiri}}\ and\ \bibinfo {author} {\bibfnamefont {S.~D.}\ \bibnamefont {Odintsov}},\ }\href {\doibase 10.1016/j.physrep.2011.04.001} {\bibfield  {journal} {\bibinfo  {journal} {Phys. Rept.}\ }\textbf {\bibinfo {volume} {505}},\ \bibinfo {pages} {59} (\bibinfo {year} {2011})}\BibitemShut {NoStop}%
\bibitem [{\citenamefont {Sotiriou}\ and\ \citenamefont {Faraoni}(2010)}]{fder1}%
  \BibitemOpen
  \bibfield  {author} {\bibinfo {author} {\bibfnamefont {T.~P.}\ \bibnamefont {Sotiriou}}\ and\ \bibinfo {author} {\bibfnamefont {V.}~\bibnamefont {Faraoni}},\ }\href {\doibase 10.1103/RevModPhys.82.451} {\bibfield  {journal} {\bibinfo  {journal} {Rev. Mod. Phys.}\ }\textbf {\bibinfo {volume} {82}},\ \bibinfo {pages} {451} (\bibinfo {year} {2010})}\BibitemShut {NoStop}%
\bibitem [{\citenamefont {Harko}\ and\ \citenamefont {Lobo}(2010{\natexlab{a}})}]{Harko2010EPJC}%
  \BibitemOpen
  \bibfield  {author} {\bibinfo {author} {\bibfnamefont {T.}~\bibnamefont {Harko}}\ and\ \bibinfo {author} {\bibfnamefont {F.~S.~N.}\ \bibnamefont {Lobo}},\ }\href {\doibase 10.1140/epjc/s10052-010-1467-3} {\bibfield  {journal} {\bibinfo  {journal} {Eur. Phys. J. C}\ }\textbf {\bibinfo {volume} {70}},\ \bibinfo {pages} {373} (\bibinfo {year} {2010}{\natexlab{a}})}\BibitemShut {NoStop}%
\bibitem [{\citenamefont {Harko}\ \emph {et~al.}(2011)\citenamefont {Harko}, \citenamefont {Lobo}, \citenamefont {Nojiri},\ and\ \citenamefont {Odintsov}}]{Harko2011PRD}%
  \BibitemOpen
  \bibfield  {author} {\bibinfo {author} {\bibfnamefont {T.}~\bibnamefont {Harko}}, \bibinfo {author} {\bibfnamefont {F.~S.~N.}\ \bibnamefont {Lobo}}, \bibinfo {author} {\bibfnamefont {S.}~\bibnamefont {Nojiri}}, \ and\ \bibinfo {author} {\bibfnamefont {S.~D.}\ \bibnamefont {Odintsov}},\ }\href {\doibase 10.1103/PhysRevD.84.024020} {\bibfield  {journal} {\bibinfo  {journal} {Phys. Rev. D}\ }\textbf {\bibinfo {volume} {84}},\ \bibinfo {pages} {024020} (\bibinfo {year} {2011})}\BibitemShut {NoStop}%
\bibitem [{\citenamefont {Harko}\ and\ \citenamefont {Lobo}(2010{\natexlab{b}})}]{Harko:2010mv}%
  \BibitemOpen
  \bibfield  {author} {\bibinfo {author} {\bibfnamefont {T.}~\bibnamefont {Harko}}\ and\ \bibinfo {author} {\bibfnamefont {F.~S.~N.}\ \bibnamefont {Lobo}},\ }\href {\doibase 10.1140/epjc/s10052-010-1467-3} {\bibfield  {journal} {\bibinfo  {journal} {Eur. Phys. J. C}\ }\textbf {\bibinfo {volume} {70}},\ \bibinfo {pages} {373} (\bibinfo {year} {2010}{\natexlab{b}})}\BibitemShut {NoStop}%
\bibitem [{\citenamefont {Odintsov}\ and\ \citenamefont {S\'aez-G\'omez}(2013)}]{Odintsov:2013iba}%
  \BibitemOpen
  \bibfield  {author} {\bibinfo {author} {\bibfnamefont {S.~D.}\ \bibnamefont {Odintsov}}\ and\ \bibinfo {author} {\bibfnamefont {D.}~\bibnamefont {S\'aez-G\'omez}},\ }\href {\doibase 10.1016/j.physletb.2013.07.026} {\bibfield  {journal} {\bibinfo  {journal} {Phys. Lett. B}\ }\textbf {\bibinfo {volume} {725}},\ \bibinfo {pages} {437} (\bibinfo {year} {2013})}\BibitemShut {NoStop}%
\bibitem [{\citenamefont {Harko}\ \emph {et~al.}(2021)\citenamefont {Harko}, \citenamefont {Myrzakulov}, \citenamefont {Myrzakulov},\ and\ \citenamefont {Shahidi}}]{HARKO2021100886}%
  \BibitemOpen
  \bibfield  {author} {\bibinfo {author} {\bibfnamefont {T.}~\bibnamefont {Harko}}, \bibinfo {author} {\bibfnamefont {N.}~\bibnamefont {Myrzakulov}}, \bibinfo {author} {\bibfnamefont {R.}~\bibnamefont {Myrzakulov}}, \ and\ \bibinfo {author} {\bibfnamefont {S.}~\bibnamefont {Shahidi}},\ }\href {\doibase https://doi.org/10.1016/j.dark.2021.100886} {\bibfield  {journal} {\bibinfo  {journal} {Phys. Dark Univ.}\ }\textbf {\bibinfo {volume} {34}},\ \bibinfo {pages} {100886} (\bibinfo {year} {2021})}\BibitemShut {NoStop}%
\bibitem [{\citenamefont {Haghani}\ and\ \citenamefont {Harko}(2021)}]{Haghani2021}%
  \BibitemOpen
  \bibfield  {author} {\bibinfo {author} {\bibfnamefont {Z.}~\bibnamefont {Haghani}}\ and\ \bibinfo {author} {\bibfnamefont {T.}~\bibnamefont {Harko}},\ }\href {\doibase 10.1140/epjc/s10052-021-09359-3} {\bibfield  {journal} {\bibinfo  {journal} {Eur. Phys. J. C}\ }\textbf {\bibinfo {volume} {81}},\ \bibinfo {pages} {615} (\bibinfo {year} {2021})}\BibitemShut {NoStop}%
\bibitem [{\citenamefont {Haghani}\ \emph {et~al.}(2013)\citenamefont {Haghani}, \citenamefont {Harko}, \citenamefont {Sepangi},\ and\ \citenamefont {Shahidi}}]{Haghani2013}%
  \BibitemOpen
  \bibfield  {author} {\bibinfo {author} {\bibfnamefont {Z.}~\bibnamefont {Haghani}}, \bibinfo {author} {\bibfnamefont {T.}~\bibnamefont {Harko}}, \bibinfo {author} {\bibfnamefont {H.~R.}\ \bibnamefont {Sepangi}}, \ and\ \bibinfo {author} {\bibfnamefont {S.}~\bibnamefont {Shahidi}},\ }\href {\doibase 10.1103/PhysRevD.88.044023} {\bibfield  {journal} {\bibinfo  {journal} {Phys. Rev. D}\ }\textbf {\bibinfo {volume} {88}},\ \bibinfo {pages} {044023} (\bibinfo {year} {2013})}\BibitemShut {NoStop}%
\bibitem [{\citenamefont {Mota}\ \emph {et~al.}(2024)\citenamefont {Mota}, \citenamefont {Pretel},\ and\ \citenamefont {Flores}}]{Mota2024}%
  \BibitemOpen
  \bibfield  {author} {\bibinfo {author} {\bibfnamefont {C.~E.}\ \bibnamefont {Mota}}, \bibinfo {author} {\bibfnamefont {J.~M.~Z.}\ \bibnamefont {Pretel}}, \ and\ \bibinfo {author} {\bibfnamefont {C.~O.~V.}\ \bibnamefont {Flores}},\ }\href {\doibase 10.1140/epjc/s10052-024-13042-8} {\bibfield  {journal} {\bibinfo  {journal} {Eur. Phys. J. C}\ }\textbf {\bibinfo {volume} {84}},\ \bibinfo {pages} {673} (\bibinfo {year} {2024})}\BibitemShut {NoStop}%
\bibitem [{\citenamefont {Pretel}(2024)}]{Pretel2024PS}%
  \BibitemOpen
  \bibfield  {author} {\bibinfo {author} {\bibfnamefont {J.~M.~Z.}\ \bibnamefont {Pretel}},\ }\href {\doibase 10.1088/1402-4896/ad5ac4} {\bibfield  {journal} {\bibinfo  {journal} {Phys. Scr.}\ }\textbf {\bibinfo {volume} {99}},\ \bibinfo {pages} {085001} (\bibinfo {year} {2024})}\BibitemShut {NoStop}%
\bibitem [{\citenamefont {Tangphati}\ \emph {et~al.}(2025)\citenamefont {Tangphati}, \citenamefont {Sakalli}, \citenamefont {Banerjee},\ and\ \citenamefont {Pradhan}}]{Tangphati2025}%
  \BibitemOpen
  \bibfield  {author} {\bibinfo {author} {\bibfnamefont {T.}~\bibnamefont {Tangphati}}, \bibinfo {author} {\bibfnamefont {I.}~\bibnamefont {Sakalli}}, \bibinfo {author} {\bibfnamefont {A.}~\bibnamefont {Banerjee}}, \ and\ \bibinfo {author} {\bibfnamefont {A.}~\bibnamefont {Pradhan}},\ }\href {\doibase 10.1088/1674-1137/ad99b2} {\bibfield  {journal} {\bibinfo  {journal} {Chinese Phys. C}\ }\textbf {\bibinfo {volume} {49}},\ \bibinfo {pages} {025110} (\bibinfo {year} {2025})}\BibitemShut {NoStop}%
\bibitem [{\citenamefont {Naseer}\ \emph {et~al.}(2025{\natexlab{a}})\citenamefont {Naseer}, \citenamefont {Sharif},\ and\ \citenamefont {Chand}}]{NASEER2025101840}%
  \BibitemOpen
  \bibfield  {author} {\bibinfo {author} {\bibfnamefont {T.}~\bibnamefont {Naseer}}, \bibinfo {author} {\bibfnamefont {M.}~\bibnamefont {Sharif}}, \ and\ \bibinfo {author} {\bibfnamefont {F.}~\bibnamefont {Chand}},\ }\href {\doibase https://doi.org/10.1016/j.dark.2025.101840} {\bibfield  {journal} {\bibinfo  {journal} {Phys. Dark Univ.}\ }\textbf {\bibinfo {volume} {48}},\ \bibinfo {pages} {101840} (\bibinfo {year} {2025}{\natexlab{a}})}\BibitemShut {NoStop}%
\bibitem [{\citenamefont {Naseer}\ \emph {et~al.}(2025{\natexlab{b}})\citenamefont {Naseer}, \citenamefont {Rehman}, \citenamefont {Sharif}, \citenamefont {Alessa},\ and\ \citenamefont {Abdel-Aty}}]{NASEER2025101958}%
  \BibitemOpen
  \bibfield  {author} {\bibinfo {author} {\bibfnamefont {T.}~\bibnamefont {Naseer}}, \bibinfo {author} {\bibfnamefont {A.}~\bibnamefont {Rehman}}, \bibinfo {author} {\bibfnamefont {M.}~\bibnamefont {Sharif}}, \bibinfo {author} {\bibfnamefont {N.}~\bibnamefont {Alessa}}, \ and\ \bibinfo {author} {\bibfnamefont {A.-H.}\ \bibnamefont {Abdel-Aty}},\ }\href {\doibase https://doi.org/10.1016/j.dark.2025.101958} {\bibfield  {journal} {\bibinfo  {journal} {Phys. Dark Univ.}\ }\textbf {\bibinfo {volume} {48}},\ \bibinfo {pages} {101958} (\bibinfo {year} {2025}{\natexlab{b}})}\BibitemShut {NoStop}%
\bibitem [{\citenamefont {Chandrasekhar}(1931)}]{Chandrasekhar:1931ih}%
  \BibitemOpen
  \bibfield  {author} {\bibinfo {author} {\bibfnamefont {S.}~\bibnamefont {Chandrasekhar}},\ }\href {\doibase 10.1086/143324} {\bibfield  {journal} {\bibinfo  {journal} {Astrophys. J.}\ }\textbf {\bibinfo {volume} {74}},\ \bibinfo {pages} {81} (\bibinfo {year} {1931})}\BibitemShut {NoStop}%
\bibitem [{\citenamefont {Chandrasekhar}(1935)}]{Chandrasekhar:1935zz}%
  \BibitemOpen
  \bibfield  {author} {\bibinfo {author} {\bibfnamefont {S.}~\bibnamefont {Chandrasekhar}},\ }\href {\doibase 10.1093/mnras/95.3.207} {\bibfield  {journal} {\bibinfo  {journal} {MNRAS}\ }\textbf {\bibinfo {volume} {95}},\ \bibinfo {pages} {207} (\bibinfo {year} {1935})}\BibitemShut {NoStop}%
\bibitem [{\citenamefont {Howell}\ \emph {et~al.}(2006)\citenamefont {Howell} \emph {et~al.}}]{SNLS:2006ics}%
  \BibitemOpen
  \bibfield  {author} {\bibinfo {author} {\bibfnamefont {D.~A.}\ \bibnamefont {Howell}} \emph {et~al.} (\bibinfo {collaboration} {SNLS}),\ }\href {\doibase 10.1038/nature05103} {\bibfield  {journal} {\bibinfo  {journal} {Nature}\ }\textbf {\bibinfo {volume} {443}},\ \bibinfo {pages} {308} (\bibinfo {year} {2006})}\BibitemShut {NoStop}%
\bibitem [{\citenamefont {Scalzo}\ \emph {et~al.}(2010)\citenamefont {Scalzo} \emph {et~al.}}]{Scalzo:2010xd}%
  \BibitemOpen
  \bibfield  {author} {\bibinfo {author} {\bibfnamefont {R.~A.}\ \bibnamefont {Scalzo}} \emph {et~al.},\ }\href {\doibase 10.1088/0004-637X/713/2/1073} {\bibfield  {journal} {\bibinfo  {journal} {\apj}\ }\textbf {\bibinfo {volume} {713}},\ \bibinfo {pages} {1073} (\bibinfo {year} {2010})}\BibitemShut {NoStop}%
\bibitem [{\citenamefont {Hicken}\ and\ \citenamefont {orhers}(2007)}]{Hicken:2007ap}%
  \BibitemOpen
  \bibfield  {author} {\bibinfo {author} {\bibfnamefont {M.}~\bibnamefont {Hicken}}\ and\ \bibinfo {author} {\bibnamefont {orhers}},\ }\href {\doibase 10.1086/523301} {\bibfield  {journal} {\bibinfo  {journal} {\apj}\ }\textbf {\bibinfo {volume} {669}},\ \bibinfo {pages} {L17} (\bibinfo {year} {2007})}\BibitemShut {NoStop}%
\bibitem [{\citenamefont {Yamanaka}\ \emph {et~al.}(2009)\citenamefont {Yamanaka} \emph {et~al.}}]{Yamanaka:2009dp}%
  \BibitemOpen
  \bibfield  {author} {\bibinfo {author} {\bibfnamefont {M.}~\bibnamefont {Yamanaka}} \emph {et~al.},\ }\href {\doibase 10.1088/0004-637X/707/2/L118} {\bibfield  {journal} {\bibinfo  {journal} {\apj}\ }\textbf {\bibinfo {volume} {707}},\ \bibinfo {pages} {L118} (\bibinfo {year} {2009})}\BibitemShut {NoStop}%
\bibitem [{\citenamefont {Silverman}\ \emph {et~al.}(2011)\citenamefont {Silverman} \emph {et~al.}}]{Silverman:2010bh}%
  \BibitemOpen
  \bibfield  {author} {\bibinfo {author} {\bibfnamefont {J.~M.}\ \bibnamefont {Silverman}} \emph {et~al.},\ }\href {\doibase 10.1111/j.1365-2966.2010.17474.x} {\bibfield  {journal} {\bibinfo  {journal} {MNRAS}\ }\textbf {\bibinfo {volume} {410}},\ \bibinfo {pages} {585} (\bibinfo {year} {2011})}\BibitemShut {NoStop}%
\bibitem [{\citenamefont {Taubenberger}\ \emph {et~al.}(2011)\citenamefont {Taubenberger} \emph {et~al.}}]{Taubenberger:2010qv}%
  \BibitemOpen
  \bibfield  {author} {\bibinfo {author} {\bibfnamefont {S.}~\bibnamefont {Taubenberger}} \emph {et~al.},\ }\href {\doibase 10.1111/j.1365-2966.2010.18107.x} {\bibfield  {journal} {\bibinfo  {journal} {MNRAS}\ }\textbf {\bibinfo {volume} {412}},\ \bibinfo {pages} {2735} (\bibinfo {year} {2011})}\BibitemShut {NoStop}%
\bibitem [{\citenamefont {Das}\ and\ \citenamefont {Mukhopadhyay}(2012)}]{Das:2012ai}%
  \BibitemOpen
  \bibfield  {author} {\bibinfo {author} {\bibfnamefont {U.}~\bibnamefont {Das}}\ and\ \bibinfo {author} {\bibfnamefont {B.}~\bibnamefont {Mukhopadhyay}},\ }\href {\doibase 10.1103/PhysRevD.86.042001} {\bibfield  {journal} {\bibinfo  {journal} {Phys. Rev. D}\ }\textbf {\bibinfo {volume} {86}},\ \bibinfo {pages} {042001} (\bibinfo {year} {2012})}\BibitemShut {NoStop}%
\bibitem [{\citenamefont {Das}\ and\ \citenamefont {Mukhopadhyay}(2013)}]{Das:2013gd}%
  \BibitemOpen
  \bibfield  {author} {\bibinfo {author} {\bibfnamefont {U.}~\bibnamefont {Das}}\ and\ \bibinfo {author} {\bibfnamefont {B.}~\bibnamefont {Mukhopadhyay}},\ }\href {\doibase 10.1103/PhysRevLett.110.071102} {\bibfield  {journal} {\bibinfo  {journal} {Phys. Rev. Lett.}\ }\textbf {\bibinfo {volume} {110}},\ \bibinfo {pages} {071102} (\bibinfo {year} {2013})}\BibitemShut {NoStop}%
\bibitem [{\citenamefont {Das}\ and\ \citenamefont {Mukhopadhyay}(2014)}]{Das:2014ssa}%
  \BibitemOpen
  \bibfield  {author} {\bibinfo {author} {\bibfnamefont {U.}~\bibnamefont {Das}}\ and\ \bibinfo {author} {\bibfnamefont {B.}~\bibnamefont {Mukhopadhyay}},\ }\href {\doibase 10.1088/1475-7516/2014/06/050} {\bibfield  {journal} {\bibinfo  {journal} {JCAP}\ }\textbf {\bibinfo {volume} {06}},\ \bibinfo {pages} {050} (\bibinfo {year} {2014})}\BibitemShut {NoStop}%
\bibitem [{\citenamefont {Franzon}\ and\ \citenamefont {Schramm}(2015)}]{Franzon:2015gda}%
  \BibitemOpen
  \bibfield  {author} {\bibinfo {author} {\bibfnamefont {B.}~\bibnamefont {Franzon}}\ and\ \bibinfo {author} {\bibfnamefont {S.}~\bibnamefont {Schramm}},\ }\href {\doibase 10.1103/PhysRevD.92.083006} {\bibfield  {journal} {\bibinfo  {journal} {Phys. Rev. D}\ }\textbf {\bibinfo {volume} {92}},\ \bibinfo {pages} {083006} (\bibinfo {year} {2015})}\BibitemShut {NoStop}%
\bibitem [{\citenamefont {Deb}\ \emph {et~al.}(2022)\citenamefont {Deb}, \citenamefont {Mukhopadhyay},\ and\ \citenamefont {Weber}}]{Deb:2021gwn}%
  \BibitemOpen
  \bibfield  {author} {\bibinfo {author} {\bibfnamefont {D.}~\bibnamefont {Deb}}, \bibinfo {author} {\bibfnamefont {B.}~\bibnamefont {Mukhopadhyay}}, \ and\ \bibinfo {author} {\bibfnamefont {F.}~\bibnamefont {Weber}},\ }\href {\doibase 10.3847/1538-4357/ac410b} {\bibfield  {journal} {\bibinfo  {journal} {Astrophys. J.}\ }\textbf {\bibinfo {volume} {926}},\ \bibinfo {pages} {66} (\bibinfo {year} {2022})}\BibitemShut {NoStop}%
\bibitem [{\citenamefont {Das}\ and\ \citenamefont {Mukhopadhyay}(2015)}]{Das:2014rca}%
  \BibitemOpen
  \bibfield  {author} {\bibinfo {author} {\bibfnamefont {U.}~\bibnamefont {Das}}\ and\ \bibinfo {author} {\bibfnamefont {B.}~\bibnamefont {Mukhopadhyay}},\ }\href {\doibase 10.1088/1475-7516/2015/05/045} {\bibfield  {journal} {\bibinfo  {journal} {JCAP}\ }\textbf {\bibinfo {volume} {05}},\ \bibinfo {pages} {045} (\bibinfo {year} {2015})}\BibitemShut {NoStop}%
\bibitem [{\citenamefont {Jing}\ and\ \citenamefont {Wen}(2016)}]{Jing_2016}%
  \BibitemOpen
  \bibfield  {author} {\bibinfo {author} {\bibfnamefont {Z.-Z.}\ \bibnamefont {Jing}}\ and\ \bibinfo {author} {\bibfnamefont {D.-H.}\ \bibnamefont {Wen}},\ }\href {\doibase 10.1088/0256-307X/33/5/050401} {\bibfield  {journal} {\bibinfo  {journal} {Chinese Phys. Lett.}\ }\textbf {\bibinfo {volume} {33}},\ \bibinfo {pages} {050401} (\bibinfo {year} {2016})}\BibitemShut {NoStop}%
\bibitem [{\citenamefont {Banerjee}\ \emph {et~al.}(2017)\citenamefont {Banerjee}, \citenamefont {Shankar},\ and\ \citenamefont {Singh}}]{Banerjee:2017uwz}%
  \BibitemOpen
  \bibfield  {author} {\bibinfo {author} {\bibfnamefont {S.}~\bibnamefont {Banerjee}}, \bibinfo {author} {\bibfnamefont {S.}~\bibnamefont {Shankar}}, \ and\ \bibinfo {author} {\bibfnamefont {T.~P.}\ \bibnamefont {Singh}},\ }\href {\doibase 10.1088/1475-7516/2017/10/004} {\bibfield  {journal} {\bibinfo  {journal} {JCAP}\ }\textbf {\bibinfo {volume} {10}},\ \bibinfo {pages} {004} (\bibinfo {year} {2017})}\BibitemShut {NoStop}%
\bibitem [{\citenamefont {Carvalho}\ \emph {et~al.}(2017)\citenamefont {Carvalho} \emph {et~al.}}]{Carvalho:2017pgk}%
  \BibitemOpen
  \bibfield  {author} {\bibinfo {author} {\bibfnamefont {G.~A.}\ \bibnamefont {Carvalho}} \emph {et~al.},\ }\href {\doibase 10.1140/epjc/s10052-017-5413-5} {\bibfield  {journal} {\bibinfo  {journal} {Eur. Phys. J. C}\ }\textbf {\bibinfo {volume} {77}},\ \bibinfo {pages} {871} (\bibinfo {year} {2017})}\BibitemShut {NoStop}%
\bibitem [{\citenamefont {Kalita}\ and\ \citenamefont {Mukhopadhyay}(2018)}]{Kalita:2018ldn}%
  \BibitemOpen
  \bibfield  {author} {\bibinfo {author} {\bibfnamefont {S.}~\bibnamefont {Kalita}}\ and\ \bibinfo {author} {\bibfnamefont {B.}~\bibnamefont {Mukhopadhyay}},\ }\href {\doibase 10.1088/1475-7516/2018/09/007} {\bibfield  {journal} {\bibinfo  {journal} {JCAP}\ }\textbf {\bibinfo {volume} {09}},\ \bibinfo {pages} {007} (\bibinfo {year} {2018})}\BibitemShut {NoStop}%
\bibitem [{\citenamefont {Eslam~Panah}\ and\ \citenamefont {Liu}(2019)}]{EslamPanah:2018evk}%
  \BibitemOpen
  \bibfield  {author} {\bibinfo {author} {\bibfnamefont {B.}~\bibnamefont {Eslam~Panah}}\ and\ \bibinfo {author} {\bibfnamefont {H.~L.}\ \bibnamefont {Liu}},\ }\href {\doibase 10.1103/PhysRevD.99.104074} {\bibfield  {journal} {\bibinfo  {journal} {Phys. Rev. D}\ }\textbf {\bibinfo {volume} {99}},\ \bibinfo {pages} {104074} (\bibinfo {year} {2019})}\BibitemShut {NoStop}%
\bibitem [{\citenamefont {Liu}\ and\ \citenamefont {L\"u}(2019)}]{Liu:2018jhd}%
  \BibitemOpen
  \bibfield  {author} {\bibinfo {author} {\bibfnamefont {H.~L.}\ \bibnamefont {Liu}}\ and\ \bibinfo {author} {\bibfnamefont {G.~L.}\ \bibnamefont {L\"u}},\ }\href {\doibase 10.1088/1475-7516/2019/02/040} {\bibfield  {journal} {\bibinfo  {journal} {JCAP}\ }\textbf {\bibinfo {volume} {02}},\ \bibinfo {pages} {040} (\bibinfo {year} {2019})}\BibitemShut {NoStop}%
\bibitem [{\citenamefont {Rocha}\ \emph {et~al.}(2020)\citenamefont {Rocha}, \citenamefont {Carvalho}, \citenamefont {Deb},\ and\ \citenamefont {Malheiro}}]{Rocha:2019nze}%
  \BibitemOpen
  \bibfield  {author} {\bibinfo {author} {\bibfnamefont {F.}~\bibnamefont {Rocha}}, \bibinfo {author} {\bibfnamefont {G.~A.}\ \bibnamefont {Carvalho}}, \bibinfo {author} {\bibfnamefont {D.}~\bibnamefont {Deb}}, \ and\ \bibinfo {author} {\bibfnamefont {M.}~\bibnamefont {Malheiro}},\ }\href {\doibase 10.1103/PhysRevD.101.104008} {\bibfield  {journal} {\bibinfo  {journal} {Phys. Rev. D}\ }\textbf {\bibinfo {volume} {101}},\ \bibinfo {pages} {104008} (\bibinfo {year} {2020})}\BibitemShut {NoStop}%
\bibitem [{\citenamefont {Wojnar}(2021)}]{Wojnar:2020ckw}%
  \BibitemOpen
  \bibfield  {author} {\bibinfo {author} {\bibfnamefont {A.}~\bibnamefont {Wojnar}},\ }\href {\doibase 10.1142/S0219887821400065} {\bibfield  {journal} {\bibinfo  {journal} {Int. J. Geom. Meth. Mod. Phys.}\ }\textbf {\bibinfo {volume} {18}},\ \bibinfo {pages} {2140006} (\bibinfo {year} {2021})}\BibitemShut {NoStop}%
\bibitem [{\citenamefont {Kalita}\ \emph {et~al.}(2021)\citenamefont {Kalita}, \citenamefont {Mukhopadhyay},\ and\ \citenamefont {Govindarajan}}]{Kalita:2019yaj}%
  \BibitemOpen
  \bibfield  {author} {\bibinfo {author} {\bibfnamefont {S.}~\bibnamefont {Kalita}}, \bibinfo {author} {\bibfnamefont {B.}~\bibnamefont {Mukhopadhyay}}, \ and\ \bibinfo {author} {\bibfnamefont {T.~R.}\ \bibnamefont {Govindarajan}},\ }\href {\doibase 10.1142/S0218271821500346} {\bibfield  {journal} {\bibinfo  {journal} {Int. J. Mod. Phys. D}\ }\textbf {\bibinfo {volume} {30}},\ \bibinfo {pages} {5} (\bibinfo {year} {2021})}\BibitemShut {NoStop}%
\bibitem [{\citenamefont {Sarmah}\ \emph {et~al.}(2022)\citenamefont {Sarmah}, \citenamefont {Kalita},\ and\ \citenamefont {Wojnar}}]{Sarmah2022PRD}%
  \BibitemOpen
  \bibfield  {author} {\bibinfo {author} {\bibfnamefont {L.}~\bibnamefont {Sarmah}}, \bibinfo {author} {\bibfnamefont {S.}~\bibnamefont {Kalita}}, \ and\ \bibinfo {author} {\bibfnamefont {A.}~\bibnamefont {Wojnar}},\ }\href {\doibase 10.1103/PhysRevD.105.024028} {\bibfield  {journal} {\bibinfo  {journal} {Phys. Rev. D}\ }\textbf {\bibinfo {volume} {105}},\ \bibinfo {pages} {024028} (\bibinfo {year} {2022})}\BibitemShut {NoStop}%
\bibitem [{\citenamefont {Kalita}\ and\ \citenamefont {Sarmah}(2022)}]{Kalita:2022tyx}%
  \BibitemOpen
  \bibfield  {author} {\bibinfo {author} {\bibfnamefont {S.}~\bibnamefont {Kalita}}\ and\ \bibinfo {author} {\bibfnamefont {L.}~\bibnamefont {Sarmah}},\ }\href {\doibase 10.1016/j.physletb.2022.136942} {\bibfield  {journal} {\bibinfo  {journal} {Phys. Lett. B}\ }\textbf {\bibinfo {volume} {827}},\ \bibinfo {pages} {136942} (\bibinfo {year} {2022})}\BibitemShut {NoStop}%
\bibitem [{\citenamefont {Kalita}\ \emph {et~al.}(2023)\citenamefont {Kalita}, \citenamefont {Sarmah},\ and\ \citenamefont {Wojnar}}]{Kalita2023}%
  \BibitemOpen
  \bibfield  {author} {\bibinfo {author} {\bibfnamefont {S.}~\bibnamefont {Kalita}}, \bibinfo {author} {\bibfnamefont {L.}~\bibnamefont {Sarmah}}, \ and\ \bibinfo {author} {\bibfnamefont {A.}~\bibnamefont {Wojnar}},\ }\href {\doibase 10.1103/PhysRevD.107.044072} {\bibfield  {journal} {\bibinfo  {journal} {Phys. Rev. D}\ }\textbf {\bibinfo {volume} {107}},\ \bibinfo {pages} {044072} (\bibinfo {year} {2023})}\BibitemShut {NoStop}%
\bibitem [{\citenamefont {Li}\ \emph {et~al.}(2024)\citenamefont {Li}, \citenamefont {Yang},\ and\ \citenamefont {Lin}}]{Li2024}%
  \BibitemOpen
  \bibfield  {author} {\bibinfo {author} {\bibfnamefont {J.}~\bibnamefont {Li}}, \bibinfo {author} {\bibfnamefont {B.}~\bibnamefont {Yang}}, \ and\ \bibinfo {author} {\bibfnamefont {W.}~\bibnamefont {Lin}},\ }\href {\doibase 10.1088/1475-7516/2024/04/081} {\bibfield  {journal} {\bibinfo  {journal} {JCAP}\ }\textbf {\bibinfo {volume} {04}},\ \bibinfo {pages} {081} (\bibinfo {year} {2024})}\BibitemShut {NoStop}%
\bibitem [{\citenamefont {Pretel}\ \emph {et~al.}(2025)\citenamefont {Pretel}, \citenamefont {Tangphati}, \citenamefont {İzzet Sakallı},\ and\ \citenamefont {Banerjee}}]{PRETEL2025}%
  \BibitemOpen
  \bibfield  {author} {\bibinfo {author} {\bibfnamefont {J.~M.~Z.}\ \bibnamefont {Pretel}}, \bibinfo {author} {\bibfnamefont {T.}~\bibnamefont {Tangphati}}, \bibinfo {author} {\bibnamefont {İzzet Sakallı}}, \ and\ \bibinfo {author} {\bibfnamefont {A.}~\bibnamefont {Banerjee}},\ }\href {\doibase https://doi.org/10.1016/j.physletb.2025.139581} {\bibfield  {journal} {\bibinfo  {journal} {Phys. Lett. B}\ }\textbf {\bibinfo {volume} {866}},\ \bibinfo {pages} {139581} (\bibinfo {year} {2025})}\BibitemShut {NoStop}%
\bibitem [{\citenamefont {Liu}\ \emph {et~al.}(2014)\citenamefont {Liu}, \citenamefont {Zhang},\ and\ \citenamefont {Wen}}]{Liu:2014jna}%
  \BibitemOpen
  \bibfield  {author} {\bibinfo {author} {\bibfnamefont {H.}~\bibnamefont {Liu}}, \bibinfo {author} {\bibfnamefont {X.}~\bibnamefont {Zhang}}, \ and\ \bibinfo {author} {\bibfnamefont {D.}~\bibnamefont {Wen}},\ }\href {\doibase 10.1103/PhysRevD.89.104043} {\bibfield  {journal} {\bibinfo  {journal} {Phys. Rev. D}\ }\textbf {\bibinfo {volume} {89}},\ \bibinfo {pages} {104043} (\bibinfo {year} {2014})}\BibitemShut {NoStop}%
\bibitem [{\citenamefont {Carvalho}\ \emph {et~al.}(2018)\citenamefont {Carvalho}, \citenamefont {Arba\~nil}, \citenamefont {Marinho},\ and\ \citenamefont {Malheiro}}]{Carvalho:2018kbt}%
  \BibitemOpen
  \bibfield  {author} {\bibinfo {author} {\bibfnamefont {G.~A.}\ \bibnamefont {Carvalho}}, \bibinfo {author} {\bibfnamefont {J.~D.~V.}\ \bibnamefont {Arba\~nil}}, \bibinfo {author} {\bibfnamefont {R.~M.}\ \bibnamefont {Marinho}}, \ and\ \bibinfo {author} {\bibfnamefont {M.}~\bibnamefont {Malheiro}},\ }\href {\doibase 10.1140/epjc/s10052-018-5901-2} {\bibfield  {journal} {\bibinfo  {journal} {Eur. Phys. J. C}\ }\textbf {\bibinfo {volume} {78}},\ \bibinfo {pages} {411} (\bibinfo {year} {2018})}\BibitemShut {NoStop}%
\bibitem [{\citenamefont {Boshkayev}\ \emph {et~al.}(2011)\citenamefont {Boshkayev}, \citenamefont {Rueda},\ and\ \citenamefont {Ruffini}}]{Boshkayev:2011aet}%
  \BibitemOpen
  \bibfield  {author} {\bibinfo {author} {\bibfnamefont {K.}~\bibnamefont {Boshkayev}}, \bibinfo {author} {\bibfnamefont {J.}~\bibnamefont {Rueda}}, \ and\ \bibinfo {author} {\bibfnamefont {R.}~\bibnamefont {Ruffini}},\ }\href {\doibase 10.1142/S0218301311040177} {\bibfield  {journal} {\bibinfo  {journal} {Int. J. Mod. Phys. E}\ }\textbf {\bibinfo {volume} {20}},\ \bibinfo {pages} {136} (\bibinfo {year} {2011})}\BibitemShut {NoStop}%
\bibitem [{\citenamefont {Boshkayev}\ \emph {et~al.}(2013)\citenamefont {Boshkayev}, \citenamefont {Rueda}, \citenamefont {Ruffini},\ and\ \citenamefont {Siutsou}}]{Boshkayev:2012bq}%
  \BibitemOpen
  \bibfield  {author} {\bibinfo {author} {\bibfnamefont {K.}~\bibnamefont {Boshkayev}}, \bibinfo {author} {\bibfnamefont {J.~A.}\ \bibnamefont {Rueda}}, \bibinfo {author} {\bibfnamefont {R.}~\bibnamefont {Ruffini}}, \ and\ \bibinfo {author} {\bibfnamefont {I.}~\bibnamefont {Siutsou}},\ }\href {\doibase 10.1088/0004-637X/762/2/117} {\bibfield  {journal} {\bibinfo  {journal} {Astrophys. J.}\ }\textbf {\bibinfo {volume} {762}},\ \bibinfo {pages} {117} (\bibinfo {year} {2013})}\BibitemShut {NoStop}%
\bibitem [{\citenamefont {Salpeter}(1961)}]{salpeter_energy_1961}%
  \BibitemOpen
  \bibfield  {author} {\bibinfo {author} {\bibfnamefont {E.~E.}\ \bibnamefont {Salpeter}},\ }\href {http://adsabs.harvard.edu/full/1961ApJ...134..669S} {\bibfield  {journal} {\bibinfo  {journal} {\apj}\ }\textbf {\bibinfo {volume} {134}},\ \bibinfo {pages} {669} (\bibinfo {year} {1961})}\BibitemShut {NoStop}%
\bibitem [{\citenamefont {Hamada}\ and\ \citenamefont {Salpeter}(1961)}]{hamada_models_1961}%
  \BibitemOpen
  \bibfield  {author} {\bibinfo {author} {\bibfnamefont {T.}~\bibnamefont {Hamada}}\ and\ \bibinfo {author} {\bibfnamefont {E.~E.}\ \bibnamefont {Salpeter}},\ }\href@noop {} {\bibfield  {journal} {\bibinfo  {journal} {\apj}\ }\textbf {\bibinfo {volume} {134}},\ \bibinfo {pages} {683} (\bibinfo {year} {1961})}\BibitemShut {NoStop}%
\bibitem [{\citenamefont {Otoniel}\ \emph {et~al.}(2019)\citenamefont {Otoniel}, \citenamefont {Franzon}, \citenamefont {Carvalho}, \citenamefont {Malheiro}, \citenamefont {Schramm},\ and\ \citenamefont {Weber}}]{Otoniel_2019}%
  \BibitemOpen
  \bibfield  {author} {\bibinfo {author} {\bibfnamefont {E.}~\bibnamefont {Otoniel}}, \bibinfo {author} {\bibfnamefont {B.}~\bibnamefont {Franzon}}, \bibinfo {author} {\bibfnamefont {G.~A.}\ \bibnamefont {Carvalho}}, \bibinfo {author} {\bibfnamefont {M.}~\bibnamefont {Malheiro}}, \bibinfo {author} {\bibfnamefont {S.}~\bibnamefont {Schramm}}, \ and\ \bibinfo {author} {\bibfnamefont {F.}~\bibnamefont {Weber}},\ }\href {\doibase 10.3847/1538-4357/ab24d1} {\bibfield  {journal} {\bibinfo  {journal} {\apj}\ }\textbf {\bibinfo {volume} {879}},\ \bibinfo {pages} {46} (\bibinfo {year} {2019})}\BibitemShut {NoStop}%
\bibitem [{\citenamefont {Wang}\ \emph {et~al.}(2012)\citenamefont {Wang} \emph {et~al.}}]{wang_ame2012_2012}%
  \BibitemOpen
  \bibfield  {author} {\bibinfo {author} {\bibfnamefont {M.}~\bibnamefont {Wang}} \emph {et~al.},\ }\href {http://iopscience.iop.org/article/10.1088/1674-1137/36/12/003/meta} {\bibfield  {journal} {\bibinfo  {journal} {Chinese Phys. C}\ }\textbf {\bibinfo {volume} {36}},\ \bibinfo {pages} {1603} (\bibinfo {year} {2012})}\BibitemShut {NoStop}%
\bibitem [{\citenamefont {Audi}\ \emph {et~al.}(2012)\citenamefont {Audi} \emph {et~al.}}]{audi_ame2012_2012}%
  \BibitemOpen
  \bibfield  {author} {\bibinfo {author} {\bibfnamefont {G.}~\bibnamefont {Audi}} \emph {et~al.},\ }\href {http://iopscience.iop.org/article/10.1088/1674-1137/36/12/002/meta} {\bibfield  {journal} {\bibinfo  {journal} {Chinese Phys. C}\ }\textbf {\bibinfo {volume} {36}},\ \bibinfo {pages} {1287} (\bibinfo {year} {2012})}\BibitemShut {NoStop}%
\bibitem [{\citenamefont {Shapiro}\ and\ \citenamefont {Teukolsky}(2008)}]{shapiro_black_2008}%
  \BibitemOpen
  \bibfield  {author} {\bibinfo {author} {\bibfnamefont {S.~L.}\ \bibnamefont {Shapiro}}\ and\ \bibinfo {author} {\bibfnamefont {S.~A.}\ \bibnamefont {Teukolsky}},\ }\href@noop {} {\emph {\bibinfo {title} {Black Holes, White Dwarfs, and Neutron Stars: The Physics of Compact Objects}}}\ (\bibinfo  {publisher} {John Wiley \& Sons},\ \bibinfo {year} {2008})\BibitemShut {NoStop}%
\bibitem [{\citenamefont {Chamel}\ \emph {et~al.}(2013)\citenamefont {Chamel}, \citenamefont {Fantina},\ and\ \citenamefont {Davis}}]{chamel_stability_2013}%
  \BibitemOpen
  \bibfield  {author} {\bibinfo {author} {\bibfnamefont {N.}~\bibnamefont {Chamel}}, \bibinfo {author} {\bibfnamefont {A.~F.}\ \bibnamefont {Fantina}}, \ and\ \bibinfo {author} {\bibfnamefont {P.~J.}\ \bibnamefont {Davis}},\ }\href {https://journals.aps.org/prd/abstract/10.1103/PhysRevD.88.081301} {\bibfield  {journal} {\bibinfo  {journal} {\prd}\ }\textbf {\bibinfo {volume} {88}},\ \bibinfo {pages} {081301(R)} (\bibinfo {year} {2013})}\BibitemShut {NoStop}%
\bibitem [{\citenamefont {Gamow}(1939)}]{gamow_physical_1939}%
  \BibitemOpen
  \bibfield  {author} {\bibinfo {author} {\bibfnamefont {G.}~\bibnamefont {Gamow}},\ }\href {https://journals.aps.org/pr/abstract/10.1103/PhysRev.55.718} {\bibfield  {journal} {\bibinfo  {journal} {PhRv}\ }\textbf {\bibinfo {volume} {55}},\ \bibinfo {pages} {718} (\bibinfo {year} {1939})}\BibitemShut {NoStop}%
\bibitem [{\citenamefont {Chamel}\ \emph {et~al.}(2014)\citenamefont {Chamel}, \citenamefont {Molter}, \citenamefont {Fantina},\ and\ \citenamefont {Arteaga}}]{chamel_maximum_2014}%
  \BibitemOpen
  \bibfield  {author} {\bibinfo {author} {\bibfnamefont {N.}~\bibnamefont {Chamel}}, \bibinfo {author} {\bibfnamefont {E.}~\bibnamefont {Molter}}, \bibinfo {author} {\bibfnamefont {A.}~\bibnamefont {Fantina}}, \ and\ \bibinfo {author} {\bibfnamefont {D.~P.}\ \bibnamefont {Arteaga}},\ }\href {https://journals.aps.org/prd/abstract/10.1103/PhysRevD.90.043002} {\bibfield  {journal} {\bibinfo  {journal} {\prd}\ }\textbf {\bibinfo {volume} {90}},\ \bibinfo {pages} {043002} (\bibinfo {year} {2014})}\BibitemShut {NoStop}%
\bibitem [{\citenamefont {Chamel}\ and\ \citenamefont {Fantina}(2015)}]{PhysRevD.92.023008}%
  \BibitemOpen
  \bibfield  {author} {\bibinfo {author} {\bibfnamefont {N.}~\bibnamefont {Chamel}}\ and\ \bibinfo {author} {\bibfnamefont {A.~F.}\ \bibnamefont {Fantina}},\ }\href {https://journals.aps.org/prd/abstract/10.1103/PhysRevD.92.023008} {\bibfield  {journal} {\bibinfo  {journal} {\prd}\ }\textbf {\bibinfo {volume} {92}},\ \bibinfo {pages} {023008} (\bibinfo {year} {2015})}\BibitemShut {NoStop}%
\bibitem [{\citenamefont {Sivia}\ and\ \citenamefont {Skilling}(2006)}]{sivia2006data}%
  \BibitemOpen
  \bibfield  {author} {\bibinfo {author} {\bibfnamefont {D.}~\bibnamefont {Sivia}}\ and\ \bibinfo {author} {\bibfnamefont {J.}~\bibnamefont {Skilling}},\ }\href {https://books.google.de/books?id=lYMSDAAAQBAJ} {\emph {\bibinfo {title} {Data Analysis: A Bayesian Tutorial}}},\ Oxford science publications\ (\bibinfo  {publisher} {OUP Oxford},\ \bibinfo {year} {2006})\BibitemShut {NoStop}%
\bibitem [{\citenamefont {{Goodman}}\ and\ \citenamefont {{Weare}}(2010)}]{goodman2010ensemble}%
  \BibitemOpen
  \bibfield  {author} {\bibinfo {author} {\bibfnamefont {J.}~\bibnamefont {{Goodman}}}\ and\ \bibinfo {author} {\bibfnamefont {J.}~\bibnamefont {{Weare}}},\ }\href {\doibase 10.2140/camcos.2010.5.65} {\bibfield  {journal} {\bibinfo  {journal} {CAMCoS}\ }\textbf {\bibinfo {volume} {5}},\ \bibinfo {pages} {65} (\bibinfo {year} {2010})}\BibitemShut {NoStop}%
\bibitem [{\citenamefont {{Foreman-Mackey}}\ \emph {et~al.}(2013)\citenamefont {{Foreman-Mackey}}, \citenamefont {{Hogg}}, \citenamefont {{Lang}},\ and\ \citenamefont {{Goodman}}}]{foreman2013emcee}%
  \BibitemOpen
  \bibfield  {author} {\bibinfo {author} {\bibfnamefont {D.}~\bibnamefont {{Foreman-Mackey}}}, \bibinfo {author} {\bibfnamefont {D.~W.}\ \bibnamefont {{Hogg}}}, \bibinfo {author} {\bibfnamefont {D.}~\bibnamefont {{Lang}}}, \ and\ \bibinfo {author} {\bibfnamefont {J.}~\bibnamefont {{Goodman}}},\ }\href {\doibase 10.1086/670067} {\bibfield  {journal} {\bibinfo  {journal} {PASP}\ }\textbf {\bibinfo {volume} {125}},\ \bibinfo {pages} {306} (\bibinfo {year} {2013})},\ \Eprint {http://arxiv.org/abs/1202.3665} {arXiv:1202.3665 [astro-ph.IM]} \BibitemShut {NoStop}%
\bibitem [{\citenamefont {Bond}\ \emph {et~al.}(2017{\natexlab{a}})\citenamefont {Bond}, \citenamefont {Schaefer}, \citenamefont {Gilliland}, \citenamefont {Holberg}, \citenamefont {Mason}, \citenamefont {Lindenblad}, \citenamefont {Seitz-McLeese}, \citenamefont {Arnett}, \citenamefont {Demarque}, \citenamefont {Spada} \emph {et~al.}}]{bond2017sirius}%
  \BibitemOpen
  \bibfield  {author} {\bibinfo {author} {\bibfnamefont {H.~E.}\ \bibnamefont {Bond}}, \bibinfo {author} {\bibfnamefont {G.~H.}\ \bibnamefont {Schaefer}}, \bibinfo {author} {\bibfnamefont {R.~L.}\ \bibnamefont {Gilliland}}, \bibinfo {author} {\bibfnamefont {J.~B.}\ \bibnamefont {Holberg}}, \bibinfo {author} {\bibfnamefont {B.~D.}\ \bibnamefont {Mason}}, \bibinfo {author} {\bibfnamefont {I.~W.}\ \bibnamefont {Lindenblad}}, \bibinfo {author} {\bibfnamefont {M.}~\bibnamefont {Seitz-McLeese}}, \bibinfo {author} {\bibfnamefont {W.~D.}\ \bibnamefont {Arnett}}, \bibinfo {author} {\bibfnamefont {P.}~\bibnamefont {Demarque}}, \bibinfo {author} {\bibfnamefont {F.}~\bibnamefont {Spada}},  \emph {et~al.},\ }\href {https://iopscience.iop.org/article/10.3847/1538-4357/aa6af8} {\bibfield  {journal} {\bibinfo  {journal} {Astrophys. J.}\ }\textbf {\bibinfo {volume} {840}},\ \bibinfo {pages} {70} (\bibinfo {year} {2017}{\natexlab{a}})}\BibitemShut {NoStop}%
\bibitem [{\citenamefont {Bond}\ \emph {et~al.}(2017{\natexlab{b}})\citenamefont {Bond}, \citenamefont {Bergeron},\ and\ \citenamefont {B{\'e}dard}}]{bond2017astrophysical}%
  \BibitemOpen
  \bibfield  {author} {\bibinfo {author} {\bibfnamefont {H.~E.}\ \bibnamefont {Bond}}, \bibinfo {author} {\bibfnamefont {P.}~\bibnamefont {Bergeron}}, \ and\ \bibinfo {author} {\bibfnamefont {A.}~\bibnamefont {B{\'e}dard}},\ }\href {https://iopscience.iop.org/article/10.3847/1538-4357/aa8a63} {\bibfield  {journal} {\bibinfo  {journal} {Astrophys. J.}\ }\textbf {\bibinfo {volume} {848}},\ \bibinfo {pages} {16} (\bibinfo {year} {2017}{\natexlab{b}})}\BibitemShut {NoStop}%
\bibitem [{\citenamefont {Bond}\ \emph {et~al.}(2015)\citenamefont {Bond}, \citenamefont {Gilliland}, \citenamefont {Schaefer}, \citenamefont {Demarque}, \citenamefont {Girard}, \citenamefont {Holberg}, \citenamefont {Gudehus}, \citenamefont {Mason}, \citenamefont {Kozhurina-Platais}, \citenamefont {Burleigh} \emph {et~al.}}]{bond2015hubble}%
  \BibitemOpen
  \bibfield  {author} {\bibinfo {author} {\bibfnamefont {H.~E.}\ \bibnamefont {Bond}}, \bibinfo {author} {\bibfnamefont {R.~L.}\ \bibnamefont {Gilliland}}, \bibinfo {author} {\bibfnamefont {G.~H.}\ \bibnamefont {Schaefer}}, \bibinfo {author} {\bibfnamefont {P.}~\bibnamefont {Demarque}}, \bibinfo {author} {\bibfnamefont {T.~M.}\ \bibnamefont {Girard}}, \bibinfo {author} {\bibfnamefont {J.~B.}\ \bibnamefont {Holberg}}, \bibinfo {author} {\bibfnamefont {D.}~\bibnamefont {Gudehus}}, \bibinfo {author} {\bibfnamefont {B.~D.}\ \bibnamefont {Mason}}, \bibinfo {author} {\bibfnamefont {V.}~\bibnamefont {Kozhurina-Platais}}, \bibinfo {author} {\bibfnamefont {M.~R.}\ \bibnamefont {Burleigh}},  \emph {et~al.},\ }\href {https://iopscience.iop.org/article/10.1088/0004-637X/813/2/106} {\bibfield  {journal} {\bibinfo  {journal} {Astrophys. J.}\ }\textbf {\bibinfo {volume} {813}},\ \bibinfo {pages} {106} (\bibinfo {year} {2015})}\BibitemShut {NoStop}%
\bibitem [{\citenamefont {Provencal}\ \emph {et~al.}(2002)\citenamefont {Provencal}, \citenamefont {Shipman}, \citenamefont {Koester}, \citenamefont {Wesemael},\ and\ \citenamefont {Bergeron}}]{provencal2002procyon}%
  \BibitemOpen
  \bibfield  {author} {\bibinfo {author} {\bibfnamefont {J.}~\bibnamefont {Provencal}}, \bibinfo {author} {\bibfnamefont {H.}~\bibnamefont {Shipman}}, \bibinfo {author} {\bibfnamefont {D.}~\bibnamefont {Koester}}, \bibinfo {author} {\bibfnamefont {F.}~\bibnamefont {Wesemael}}, \ and\ \bibinfo {author} {\bibfnamefont {P.}~\bibnamefont {Bergeron}},\ }\href {https://iopscience.iop.org/article/10.1086/338769} {\bibfield  {journal} {\bibinfo  {journal} {Astrophys. J.}\ }\textbf {\bibinfo {volume} {568}},\ \bibinfo {pages} {324} (\bibinfo {year} {2002})}\BibitemShut {NoStop}%
\bibitem [{\citenamefont {Caiazzo}\ \emph {et~al.}(2021)\citenamefont {Caiazzo}, \citenamefont {Burdge}, \citenamefont {Fuller}, \citenamefont {Heyl}, \citenamefont {Kulkarni}, \citenamefont {Prince}, \citenamefont {Richer}, \citenamefont {Schwab}, \citenamefont {Andreoni}, \citenamefont {Bellm} \emph {et~al.}}]{caiazzo2021highly}%
  \BibitemOpen
  \bibfield  {author} {\bibinfo {author} {\bibfnamefont {I.}~\bibnamefont {Caiazzo}}, \bibinfo {author} {\bibfnamefont {K.~B.}\ \bibnamefont {Burdge}}, \bibinfo {author} {\bibfnamefont {J.}~\bibnamefont {Fuller}}, \bibinfo {author} {\bibfnamefont {J.}~\bibnamefont {Heyl}}, \bibinfo {author} {\bibfnamefont {S.}~\bibnamefont {Kulkarni}}, \bibinfo {author} {\bibfnamefont {T.~A.}\ \bibnamefont {Prince}}, \bibinfo {author} {\bibfnamefont {H.~B.}\ \bibnamefont {Richer}}, \bibinfo {author} {\bibfnamefont {J.}~\bibnamefont {Schwab}}, \bibinfo {author} {\bibfnamefont {I.}~\bibnamefont {Andreoni}}, \bibinfo {author} {\bibfnamefont {E.~C.}\ \bibnamefont {Bellm}},  \emph {et~al.},\ }\href {https://www.nature.com/articles/s41586-021-03615-y} {\bibfield  {journal} {\bibinfo  {journal} {Nat.}\ }\textbf {\bibinfo {volume} {595}},\ \bibinfo {pages} {39} (\bibinfo {year} {2021})}\BibitemShut {NoStop}%
\bibitem [{\citenamefont {Potekhin}\ \emph {et~al.}(1999)\citenamefont {Potekhin}, \citenamefont {Baiko}, \citenamefont {Haensel},\ and\ \citenamefont {Yakovlev}}]{1999A&A...346..345P}%
  \BibitemOpen
  \bibfield  {author} {\bibinfo {author} {\bibfnamefont {A.~Y.}\ \bibnamefont {Potekhin}}, \bibinfo {author} {\bibfnamefont {D.}~\bibnamefont {Baiko}}, \bibinfo {author} {\bibfnamefont {P.}~\bibnamefont {Haensel}}, \ and\ \bibinfo {author} {\bibfnamefont {D.~G.}\ \bibnamefont {Yakovlev}},\ }\href {\doibase 10.48550/arXiv.astro-ph/9903127} {\bibfield  {journal} {\bibinfo  {journal} {A\&A}\ }\textbf {\bibinfo {volume} {346}},\ \bibinfo {pages} {345} (\bibinfo {year} {1999})}\BibitemShut {NoStop}%
\end{thebibliography}
\end{document}